\documentclass[a4paper,12pt]{article}   

\usepackage{amsmath,amsthm,amssymb,cite,enumerate,rotating,graphicx} 

\usepackage[usenames,dvipsnames]{color}	

\def \BEA { \begin{eqnarray}}
\def \EEA {\end{eqnarray}}
\def \BE {\begin{equation}}
\def \EE {\end{equation}}

\newcommand{\WDS}[1] {\Phi_{#1}^{S}}
\newcommand{\WDA}[1] {\Phi_{#1}^{A}}
\newcommand{\WD}[1] {\Phi_{#1}}

\newcommand{\OO}[1] {{O}(r^{-#1})}
\newcommand{\OOp}[1] {{O}(r^{#1})}
\newcommand{\OOn} {{O}(1)}
\newcommand{\oo}[1] {o(r^{-#1})}
\newcommand{\oop}[1] {o(r^{#1})}
\newcommand{\oon} {{o}(1)}

\newcommand{\R}[1] {\rho_{#1}}		

\newcommand{\Om}[1] {\Omega_{#1}}	
\newcommand{\Ps}[1] {\Psi_{#1}}

\newcommand{\om}[1] {\omega_{#1}}	

\newcommand{\ta}[1] {L_{#11}}	
\newcommand{\N}[1] {N_{#1}}

\newcommand{\Rc}[1] {(11#1,\cite{OrtPraPra07})}		

\newcommand{\Bi}[1] {(B#1,\cite{Pravdaetal04})}		

\def \mi {\stackrel{i}{m}}
\def \mj {\stackrel{j}{m}}
\def \mk {\stackrel{k}{m}}
\def \mr {\stackrel{r}{m}}
\def \ms {\stackrel{s}{m}}
\def \mz {\stackrel{z}{m}}
\def \mq {\stackrel{q}{m}}
\def \mo {\stackrel{o}{m}}
\def \mD {\stackrel{2}{m}}
\def \mT {\stackrel{3}{m}}
\def \mC {\stackrel{4}{m}}

\def \mio #1 {\mi_{#1}\ ^{  \! \! \! \! 0}} 
\def \mjo #1 {\mj_{#1}\ ^{  \! \! \! \! 0}} 
\def \mko #1 {\mk_{#1}\ ^{  \! \! \! \! 0}} 
\def \mro #1 {\mr_{#1}\ ^{  \! \! \! \! 0}} 
\def \mso #1 {\ms_{#1}\ ^{  \! \! \! \! 0}} 
\def \mpo #1 {\mp_{#1}\ ^{  \! \! \! \! 0}} 
\def \mzo #1 {\mz_{#1}\ ^{  \! \! \! \! 0}} 
\def \mqo #1 {\mq_{#1}\ ^{  \! \! \! \! 0}} 
\def \moo #1 {\mo_{#1}\ ^{  \! \! \! \! 0}} 
\def \mDo #1 {\mD_{#1}\ ^{  \! \! \! \! 0}} 
\def \mTo #1 {\mT_{#1}\ ^{  \! \! \! \! 0}} 
\def \mCo #1 {\mC_{#1}\ ^{  \! \! \! \! 0}} 

\def \miJ #1 {\mi_{#1}\ ^{  \! \! \! \! (1)}} 
\def \mjJ #1 {\mj_{#1}\ ^{  \! \! \! \! (1)}} 
\def \mkJ #1 {\mk_{#1}\ ^{  \! \! \! \! (1)}} 
\def \mrJ #1 {\mr_{#1}\ ^{  \! \! \! \! (1)}}

\def \bl {\mbox{\boldmath{$\ell$}}}
\def \hbl {\mbox{\boldmath{$\hat \ell$}}}
\def \bn {\mbox{\boldmath{$n$}}}

\def \hbn {\mbox{\boldmath{$\hat n$}}}
\def \hbm #1 {\mbox{\boldmath{$\hat m^{(#1)}$}}}
\def \bm {\mbox{\boldmath{$m$}}}

\newcommand{\be}{\begin{equation}}
\newcommand{\ee}{\end{equation}}
\newcommand{\beqn}{\begin{eqnarray}}
\newcommand{\eeqn}{\end{eqnarray}}
\newcommand{\pa}{\partial}
\newcommand{\ba}{\begin{array}}
\newcommand{\ea}{\end{array}}

\def \BEAH {\begin{eqnarray*}}
\def \EEAH {\end{eqnarray*}}
\def \BEA {\begin{eqnarray}}
\def \EEA {\end{eqnarray}}
\def \BDM {\begin{displaymath}}
\def \EDM {\end{displaymath}}

\def \pul {{{\footnotesize{\frac{1}{2}}}}}

\newcommand{\M}[3] {{\stackrel{#1}{M}}_{{#2}{#3}}}
\newcommand{\m}[3] {{\stackrel{\hspace{.3cm}#1}{m}}_{\!{#2}{#3}}\,}

\begin{document}

\title{Asymptotic behaviour of the Weyl tensor in higher dimensions}

\author{Marcello Ortaggio\thanks{ortaggio@math.cas.cz}, Alena Pravdov\' a\thanks{pravdova@math.cas.cz} \\
Institute of Mathematics, Academy of Sciences of the Czech Republic \\ \v Zitn\' a 25, 115 67 Prague 1, Czech Republic}
\date{\today}

\maketitle

\begin{abstract}

We determine the leading order fall-off behaviour of the Weyl tensor in higher dimensional Einstein spacetimes (with and without a cosmological constant) as one approaches infinity along a congruence of null geodesics. The null congruence is assumed to ``expand'' in all directions near infinity (but it is otherwise generic), which includes in particular asymptotically flat spacetimes. In contrast to the well-known four-dimensional peeling property, the fall-off rate of various Weyl components depends substantially on the chosen boundary conditions, and is also influenced by the presence of a cosmological constant. The leading component is always algebraically special, but in various cases it can be of type N, III or II.

\end{abstract}


\tableofcontents

\section{Introduction}

\label{sec_introd}

The study of isolated systems in general relativity is based on the analysis of asymptotic properties of spacetimes. Under certain assumptions, this enables one to define physical quantities such as mass, angular momentum and energy flux. In particular, properties of gravitational radiation can be determined by considering the spacetime behaviour ``far away'' along a geodesic null congruence.

In four dimensions, the Weyl tensor decay is described by the well-known peeling property, i.e., components of boost weight (b.w.) $w$ fall off as $1/r^{w+3}$ (where $w=\pm 2, \pm1,0$, and the $1/r$ term characterizes radiative fields). This result was obtained by coordinate-based approaches that studied Einstein's vacuum equations {assuming}  suitable asymptotic ``outgoing radiation'' conditions, which were formulated in terms either of the metric coefficients \cite{BBM,Sachs62} or directly of the Weyl tensor \cite{NP,NewUnt62} (see \cite{RobTra60,sachspropagation,Sachs61} for early results in special cases). 
From a more geometrical viewpoint, the peeling-off behaviour also naturally follows from Penrose's conformal definition of asymptotically simple spacetimes (which also allows for a cosmological constant) \cite{Penrose63,Penrose65prs}, at least under suitable smoothness conditions on the conformal geometry (see also \cite{penrosebook2}).

In an $n$-dimensional spacetime, the definition of asymptotic flatness at null infinity ({along with} the ``news'' tensor and Bondi energy-momentum) using a conformal method  turns out to be sound only for even $n$ \cite{HolIsh05} (see also \cite{Ishibashi08}) -- linear gravitational perturbation of the metric tensor typically decays as $r^{-(n/2-1)}$ and the unphysical (conformal) metric is thus not smooth at null infinity {if $n$ is odd (see \cite{ChoChrLoi06} for further results for even $n$)}. In \cite{HolWal04}, linear (vacuum) perturbations of Minkowski spacetime were studied in terms of the Weyl tensor, which was found to decay as $r^{-(n/2-1)}$, thus again non-smoothly in odd dimensions.\footnote{In the present paper we discuss the {\em physical} Weyl tensor only, so here we have accordingly rephrased the results of \cite{HolWal04} (where the {\em unphysical} Weyl tensor {of the conformal spacetime} was instead considered).} Ref.~\cite{HolWal04} also pointed out a qualitative difference between $n=4$ and $n>4$ in the decay properties of various Weyl components at null infinity and related this to a possible new peeling behaviour when $n>4$. This expectation was indeed confirmed {in the full theory} in \cite{GodRea12} by {studying} the Bondi-like metric defined in \cite{TanTanShi10,TanKinShi11} (also mentioned in \cite{HolIsh05,Ishibashi08}) and thus an expansion {of the Weyl tensor} along the generators of a family of outgoing null hypersurfaces. Not only was the $r^{-(n/2-1)}$ result of \cite{HolWal04} recovered at the leading order, but at higher orders a new structure of the $r$-dependence of various Weyl components was also obtained \cite{GodRea12}. For odd $n$, an extra condition on the asymptotic metric coefficients {was} needed in \cite{GodRea12} (see also \cite{TanTanShi10}), in relation to the {simultaneous} appearance of {integer and} semi-integer powers in the expansions. (Note that the analysis of \cite{GodRea12} includes not only vacuum spacetimes but also possible matter fields that decay ``fast enough'' at infinity, cf.~\cite{GodRea12} for details.)

The present contribution studies the asymptotic behaviour of the Weyl tensor {in higher  dimensional {\em Einstein spacetimes} ($R_{ab}=\frac{R}{n}g_{ab}$) under more general boundary conditions, for which a different method seems to be more suitable. The basic idea is still to} evaluate the Weyl components 
in a frame parallelly transported along a congruence of ``outgoing'' null geodesics, {affinely parametrized by $r$ (the congruence is rather ``generic'' and not assumed to be hypersurface orthogonal} -- its precise properties will be specified in section~\ref{subsec_sachs} below). However, {on the lines of the classic 4D work \cite{NP},} we do not make assumptions on the spacetime metric but work directly with the Weyl tensor, in the framework of the higher dimensional Newman-Penrose (NP) formalism \cite{Pravdaetal04,Coleyetal04vsi,OrtPraPra07,Durkeeetal10,OrtPraPra12rev} (we follow the notation of {the review} \cite{OrtPraPra12rev} and we do not repeat here the definitions of all the symbols). 
{This} permits a unified study for both even and odd dimensions, and with little extra effort it also allows for a possible cosmological constant.  In the case of asymptotically flat spacetimes {the Bianchi equations naturally give} the ``$r^{-(n/2-1)}$-result'' for the leading Weyl components (see \eqref{AF} below), {as previously obtained with the methods of \cite{HolWal04,GodRea12}}. 
In addition to this special case, a complete pattern of possible fall-off behaviours both with (sections \ref{subsubsec_R_summary}, \ref{subsec_R_ii}, \ref{subsec_R_iii}) and without (sections \ref{subsubsec_R=0_summary}, \ref{subsec_R=0_ii}, \ref{subsec_R=0_iii}) a cosmological constant is presented. The precise fall-off for a specific spacetime will be determined by a choice of ``boundary condition'' at null infinity. These are naturally specified by first fixing a bound on the decay rate of b.w. +2 Weyl components $\Om{ij}$ (which we will assume to be faster than $1/r^2$), as in four dimensions. However, while in 4D only the fall-off $\Om{ij}=\OO{5}$ needs to be assumed (and then the standard peeling result follows \cite{NP}),\footnote{The $\Om{ij}$ components of the $n$-dimensional notation correspond to the NP scalar $\Psi_0$ in 4D.} for $n>4$ the $r$-dependence of the remaining Weyl components will still be partially undetermined and various possible choices of boundary conditions for lower b.w. components will lead to different fall-off behaviours. More specifically, how such numerous cases (and subcases) arise can be better understood by observing that the Weyl components containing arbitrary integration ``constants'' are $\Ps{ijk}$ (at order $1/r^n$ or $1/r^3$) and, for $n>5$, $\WD{ijkl}$  (at order $1/r^2$). This will be worked out in the paper.\footnote{To be precise, by ``arbitrary integration constants'' we refer to $r$-independent quantities that generically may still depend on coordinates {different from $r$.} Additionally, (some of) these may be ``arbitrary'' only at the level of the $r$-integration of the (asymptotic) NP equations -- the remaining ``transverse'' NP equations would in fact play a role of ``constraint equations''. This is of course important for a full analysis of the characteristic initial value problem, but it goes beyond the scope of this paper and will not be discussed in the following (for details in 4D see \cite{NewUnt62} and, e.g., the review \cite{NewTod80}).} 

{Certain cases of physical interest (including asymptotically (A)dS and asymptotically flat spacetime) arise when we set to zero the terms of order $1/r^3$ in  $\Ps{ijk}$ and $1/r^2$ in $\WD{ijkl}$.
For $R\neq0$, we then obtain that necessarily $\Om{ij}=\OO{1-n}$ (or faster), and the fall-off generically is (see \eqref{R_ii_betac<-2}) 
\beqn
	& & \Om{ij}=\OO{1-n} , \nonumber \\
	& & \Ps{ijk}=\OO{n} , \nonumber \\
	& & \WD{ijkl}=\OOp{1-n}, \qquad \qquad \WDA{ij}={\OO{n}}   \qquad (R\neq0) , \label{AdS} \\
	& & \Ps{ijk}'={\OOp{2-n}}    ,\nonumber \\
	& & {\Om{ij}'=\OOp{3-n}} , \nonumber 
\eeqn
where components are ordered by decreasing b.w.. Under the same assumptions, more possibilities arise for a vanishing cosmological constant, depending more substantially on the precise fall-off prescribed for $\Om{ij}$. In particular, if $\Om{ij}$ falls faster than $1/r^{n/2}$ but not faster than $1/r^{n/2+1}$ we have (cf.~\eqref{R=0_i_beta=-n/2} and the discussion after it)
\beqn
			& & \Om{ij}=\OO{\nu}  \qquad \left(\frac{n}{2}<\nu\le1+\frac{n}{2}\right) , \nonumber \\
			& & \Ps{ijk}=\OO{\nu} , \nonumber \\
			& & \WD{ijkl}=\OO{n/2} , \qquad \WD{}=\OO{\nu} , \qquad \WDA{ij}=\OO{\nu}  \qquad (R=0) , \label{AF} \\
			& & \Ps{ijk}'=\OO{n/2} , \nonumber  \\
			& & \Om{ij}'=\OOp{1-n/2} . \nonumber 
\eeqn	
This includes the behaviour found in \cite{GodRea12} for asymptotically flat radiative spacetimes. The radiative term {$\OOp{1-n/2}$ in $\Om{ij}'$} vanishes if $\nu>1+\frac{n}{2}$, in which case the fall-off is completely different (e.g., it is given by \eqref{R=0_ii_beta_c=1-n} for $\nu>n$, but other cases are also possible, see section \ref{sec_R=0} for details). On the other hand, if $\Om{ij}$ falls as $1/r^{n/2}$ or slower, one finds instead the behaviour \eqref{R=0_generic} (with $\nu>3$). Both \eqref{AdS} and \eqref{AF} are qualitatively different from the corresponding results \eqref{R_ii_n=4} and \eqref{R=0_n=4_ii} for the 4D case (apart from $\WDA{ij}$, \eqref{AdS} with $n=4$ would look the same as \eqref{R_ii_n=4}, but see comments in the following sections).}

{More general asymptotia can also be of physical interest and the corresponding fall-off properties are given in the paper. Let us just mention here, for example, that a non-zero term of order $1/r^2$ in $\WD{ijkl}$ may correspond, e.g., to black holes living in generic Einstein spacetimes (this is manifest in the case of static black holes from the Weyl $r$-dependence given in \cite{PodOrt06}). Although here we restrict to Einstein spacetimes, several results can presumably be easily extended to include matter fields that fall-off ``sufficiently'' fast (cf.~\cite{GodRea12}). The method employed here can also be similarly applied to more general contexts such as the coupled Einstein-Maxwell equations, which we leave for future work. We further note that previous results concerning the (exact) $r$-dependence of the Weyl tensor for algebraically special Einstein spacetimes include \cite{PodOrt06,PraPra08,OrtPraPra09,OrtPraPra09b,OrtPraPra10,MalPra11}.

\subsection*{On the invariance of the results}

Chosen a null direction $\bl$, the results we will present hold in a ``generic'' parallelly transported frame. One may thus wonder if the behaviour we find is frame-dependent. Similarly as in four dimensions, the answer follows from transformation properties of various Weyl components under null rotations about $\bl$, i.e.,
\be
 \hbl=\bl, \qquad \hbn =\bn+z_i\bm_{i} -\pul z_iz^i\bl , \qquad \mbox{\boldmath{$\hat{m}$}}_{i} =\bm_{i} -z_i\bl .
    \label{nullrot}
\ee
Two different parallelly transported frames are related by a transformation \eqref{nullrot} (apart from trivial spatial rotations) with the parameters $z_i$ being $r$-independent \cite{OrtPraPra07}. Under \eqref{nullrot}, the change of a Weyl component of a given b.w. $w$ is simply a term linear in components of b.w. {\em smaller} than $w$, with coefficients determined by the $z_i$ (see, e.g., eqs. (2.27)--(2.35) of \cite{Durkeeetal10}). It thus follows, in particular, that at the leading order (when $r\to\infty$) a certain Weyl component will be unchanged {\em if} all Weyl components of lower boost weight decay faster. This is always the case, for instance, for the b.w. -2 components $\Om{ij}'$ when the leading order term is of type N. Therefore, this observation will apply to several of the results of this paper, most notably to the radiative behaviour \eqref{AF} (or \eqref{R=0_i_beta=-n/2}), in which case the leading Weyl component can be related to the Bondi flux \cite{GodRea12}.
By contrast, when {leading}-order terms are not invariant in the sense just discussed, a transformation \eqref{nullrot} can be used to pick up preferred frames, which may simplify certain expressions and be useful for particular applications (see, e.g., {\cite{PraPra08,OrtPraPra_prep}} in the case of algebraically special spacetimes). This freedom will not be used here since we are interested in the asymptotic behaviour in a generic parallelly transported frame.

\subsection*{Assumptions and notation}

 In this paper, we are interested in determining the {\em leading}-order $r$-dependence of the Weyl tensor of Einstein spacetimes, while a {systematic} study of subleading terms and the analysis of asymptotic solutions of the NP equations is left for future work (several results have been already obtained in the case of algebraically special spacetimes \cite{OrtPraPra_prep}). For this reason we will not need to assume that the NP quanties (Weyl tensor, Ricci rotation coefficients, derivative operators) admit a series expansion. However, we will assume that for large $r$ the leading terms of those quantities have a power-like behaviour (so that for our purposes the notation $f=\OO{\zeta}$ will effectively mean $f\sim r^{-\zeta}$), where the powers will not be restricted to be integer numbers. We will also assume that if $f=\OO{\zeta}$ then $\pa_rf=\OO{\zeta-1}$ and $\pa_Af=\OO{\zeta}$ (where $\pa_A$ denote a derivative w.r.t. coordinates $x^A$ different from $r$ and that need not be further specified for our purposes). In a few cases it will be useful to consider subleading terms of some expressions (most importantly \eqref{nu>3}), and it will be understood that those are also assumed to be power-like.

Although we are not interested in giving the full set of asymptotic field equations, in some cases it will be useful to display relations among the leading terms of certain Weyl components. For a generic {frame} Weyl component ``$f$'' we thus define the notation
\be
	f=\frac{f^{(\zeta)}}{r^\zeta}+\oo{\zeta} ,
\ee
{where $f^{(\zeta)}$ does not depend on $r$} (so that we will have, e.g., $\WDS{ij}=\Phi^{S(n-1)}_{ij}r^{1-n}+\oop{1-n}$, or  $\Ps{ijk}=\Ps{ijk}^{(3)}r^{-3}+\oo{3}$, etc.). {For the Ricci rotation coefficients we will instead denote $r$-independent quantities by lowercase latin letters, e.g., $L_{1i}=l_{1i}r^{-1}+\oo{1}$, $\M{i}{j}{1}=\m{i}{j}{1}+\oon$, etc..}

Many of the equations will take a more compact form using the rescaled Ricci scalar 
\be 
 \tilde R=\frac{R}{n(n-1)}. 
\ee

We will be interested in the asymptotic behaviour along a geodesic null congruence with an affine parameter $r$ and tangent vector field $\bl$. Calculations will be performed in a frame $(\bl,\bn,\bm_{i})$ (with $i,j,k,\ldots=2,\ldots,n-1)$ which is parallelly transported along $\bl$. The above assumptions imply the vanishing of the following Ricci rotation coefficients (cf.~\cite{OrtPraPra12rev} for more details on the notation)
\be
	\kappa_i=0=L_{10} , \qquad \M{i}{j}{0}=0, \qquad N_{i0}=0 .
	\label{pt}
\ee
Directional derivatives along the frame vectors $(\bl,\bn,\bm_{i})$ will be denoted, respectively, by $D$, $\delta_i$, and $\Delta$.

Section~\ref{subsec_sachs} and the first parts of sections \ref{subsubsec_R} and \ref{subsubsec_R=0} are devoted to results on the Ricci rotation coefficients, to preliminary analysis of the Weyl tensor and to setting up the method. Readers not interested in those details can jump to the summary of the results for the Weyl tensor in sections \ref{subsubsec_R_summary}, \ref{subsec_R_ii}, \ref{subsec_R_iii} ($\tilde R\neq0$) and \ref{subsubsec_R=0_summary}, \ref{subsec_R=0_ii}, \ref{subsec_R=0_iii} ($\tilde R=0$). For comparison, four-dimensional results are also reproduced in the various cases and given in \eqref{R_i_n=4}, \eqref{R_ii_n=4} ($\tilde R\neq0$) and \eqref{R=0_i_n=4}, \eqref{R=0_n=4_nu>2}, \eqref{R=0_n=4_ii} ($\tilde R=0$).

\section{{Boundary conditions and Ricci rotation coefficients}}

{In this section, we explain our assumptions on the asymptotic behaviour of $\bl$ and of the Weyl tensor components of b.w. +2, and use those to fix the leading-order behaviour of the Ricci rotation coefficients and derivative operators (both for $\tilde R\neq0$ and $\tilde R=0$). It will also follow that subsequent analysis will need to consider three different choices of boundary conditions on the Weyl components of b.w. +1, which we will do in later sections.} 

\subsection{Sachs equation and optical matrix}

\label{subsec_sachs}

{In the frame $(\bl,\bn,\bm_{i})$ (see above), the optical matrix of $\bl=\pa_r$ is given by
\be
 \R{ij}=\ell_{a;b}m^a_{(i)}m^b_{(j)} .
\label{optical}
\ee
From now on,} we assume that $\R{ij}$ is {\em asymptotically non-singular {and expanding}}, i.e., the leading term of $\R{ij}$ (for large $r$) is a matrix with non-zero determinant {and non-zero trace. Roughly speaking, this means that near infinity $\bl$ expands in all spacelike directions at the same speed, which is compatible, in particular, with asymptotically flat spacetimes (as follows from \cite{TanTanShi10,TanKinShi11,GodRea12} -- however we will see in the following that this assumptions hold also in more general spacetimes).}

{Next, one needs to specify the speed at which the Weyl tensor tends to zero for $r\to\infty$. In general, we will make only the following rather weak assumption for the} fall-off for the b.w. +2 components of the Weyl tensor 
\be
	\Om{ij}=\OO{\nu} ,  \qquad {\nu>2} , 
	\label{Omega}
\ee	
although, in most cases of interest, $\nu$ will in fact be larger, as we will show {(recall that in four dimensions the existence of a smooth null infinity requires $\nu\ge5$ \cite{NP,Penrose63,Penrose65prs,penrosebook2}).}

With the assumptions listed above {the Sachs equation reads $D\R{ij}=-\R{ik}\R{kj}-\Om{ij}$ (cf.~\Rc{g}), from which one finds}\footnote{{Another solution is $\R{ij}=\OO{\nu+1}$ (for $\nu>2$), which however gives an asymptotically non-expanding optical matrix (since $\Om{ij}$ is traceless), contrary to our assumptions.}} 

\be
 \R{ij}=\frac{\delta_{ij}}{r}+\oo{1} .
\label{L_2}
\ee

In general, it is easy to see from \Rc{g} that {\em $\Om{ij}$ will affect $\R{ij}$ at order $\OO{\nu+1}$}. At all {lower} orders, the $r$-dependence of $\R{ij}$ is given by negative integer powers of $r$, which can be fixed recursively as done (to arbitrary order) in \cite{OrtPraPra_prep}. Thus, for example, if $\nu>3$ (which will indeed occur in several cases discussed in the following) one has 
\be
 \R{ij}=\frac{\delta_{ij}}{r}+\frac{b_{ij}}{r^2}+\oo{2} \qquad (\nu>3) , 
\label{nu>3}
\ee
where the subleading term contains an arbitrary ``integration matrix'' $b_{ij}$ {independent of $r$}. Note that when $\bl$ is twistfree then $b_{[ij]}=0$ (the reverse is also true if $\bl$ is a WAND \cite{OrtPraPra09b}).

{Since we have now outlined all our assumptions (see also section~\ref{sec_introd}), for readers' convenience let us summarize those before proceeding: (i) the spacetimes in question are Einstein (possibly, Ricci flat); (ii)  $\bl=\pa_r$ is a vector field tangent to a congruence of null geodesics, affinely parametrized by $r$; (iii) a frame $(\bl,\bn,\bm_{i})$ parallelly transported along $\bl$ is employed (so that \eqref{pt} holds); (iv) the optical matrix of $\bl$ is asymptically non-singular and expanding (as defined by \eqref{optical} and the following comments); (v) near infinity (i.e., $r\to\infty$) the frame components of the Weyl tensor, of the Ricci rotations coefficients, and the derivative operators admit a power-like behaviour at the leading order (in very few cases also at the subleading order, as explained in the text); (vi) the b.w. +2 components of the Weyl tensor fall off as $\Om{ij}=\OO{\nu}$, with $\nu>2$ (eq.~\eqref{Omega}). More specific possible choices of values (or range of values) of $\nu$ will determine various fall-off patterns of the remaining Weyl components, as explained in the following sections and summarized in final tables~\ref{tab_R} and \ref{tab_R=0}. We further observe that (again depending on $\nu$) in certain cases it will later be necessary also to specify the fall-off of the b.w. +1 components $\Ps{ijk}$ (see section~\ref{subsec_Rc0} below) and the b.w. 0 components $\WD{ijkl}$ -- all possible cases will be considered, and again we refer to tables~\ref{tab_R} and \ref{tab_R=0} for a summary of those.}

\subsection{Derivative operators and commutators}

\label{subsec_commut}

Taking $r$ as one of the coordinates we can write 
\be
 D=\pa_r , \qquad \Delta=U\pa_r+X^A\pa_A , \qquad \delta_i=\omega_i\pa_r+\xi^A_i\pa_A ,
\label{derivatives}
\ee
where $\pa_A=\pa/\pa x^A$ and the $x^A$ represent any set of ($n-1$) scalar functions such that $(r,x^A)$ is a well-behaved coordinate system {(at least locally near infinity, which suffices for our purposes)}. From the commutators \cite{Coleyetal04vsi} 
\beqn
 & & \Delta D-D\Delta=L_{11}D+L_{i1}\delta_i , \label{comm_DelD} \\
 & & \delta_i D-D\delta_i=L_{1i}D+\R{ji}\delta_j , \label{comm_dD}
\eeqn
we obtain the differential equations (cf. also \cite{OrtPraPra_prep})
\beqn
 & & D\omega_i=-L_{1i}-\R{ji}\omega_j , \label{comm_om} \\
 & & D\xi^A_i=-\R{ji}\xi^A_j , \label{comm_xi} \\
 & & DU=-L_{11}-L_{i1}\omega_i, \label{comm_U} \\
 & & DX^A=-L_{i1}\xi^A_i  \label{comm_X} .
\eeqn

Using \eqref{L_2}, eq.~\eqref{comm_xi} gives
\be
 \xi^A_i=\OO{1} .
\label{xi}
\ee
Similarly as mentioned above for $\R{ij}$,  $\Om{ij}$ will affect $\xi^A_i$ at order $\OO{\nu+1}$.

To fix the full $r$-dependence of the derivative operators we also need to study the behaviour of the Ricci rotation coefficients of b.w. 0 and -1. However, the corresponding differential equations will in turn involve also Weyl components of b.w. +1 and 0, respectively, and thus  one has to consider the set of the ``$D$''-Ricci identities of b.w. $b$ simultaneously with the ``$D$''-Bianchi identities of b.w. $(b+1)$ (for $b$=+1,0,-1,-2).

\subsection{Ricci rotation coefficients of b.w. 0 and Weyl components of b.w. +1}

\label{subsec_Rc0}

We need to study \Rc{b}, \Rc{e}, \Rc{n} and \Bi{8}, along with \eqref{comm_om}, \eqref{comm_X}. One starts by assuming a generic behaviour for large $r$ for each of the ``unknowns'' (e.g., $L_{1i}=\OOp{\alpha}$, where $\alpha$ need not be specified a priori). By combining conditions coming from all the considered equations one can constraint such leading terms. For example, from \Rc{b} it is easy to see that one can only have either
\be
	L_{1i}=\OO{1}, \qquad \Ps{i}=\oo{2} ,
\ee
or
\be
 L_{1i}=\OOp{\alpha}, \qquad \Ps{i}=\OOp{\alpha-1} \qquad (\alpha\neq-1) .	
\ee

Working out similar conditions for other quantities from \Rc{n}, \Bi{8} and \eqref{comm_om} and requiring compatibility of all such conditions one concludes that 
\be
	L_{1i}=\OO{1}, \qquad \M{i}{j}{k}=\OO{1} , \qquad  \om{i}=\OOn , \label{Rc0}
\ee
where it is understood that for $r\to\infty$ all terms can go to zero faster than indicated, in special cases. {However, we will  consider only the generic case, in which this does not happen.} For the Weyl tensor components of positive b.w. there are three possibilities:
\begin{enumerate}[i)]

	\item $\Ps{ijk}=\OO{\nu} , \qquad \Om{ij}=\OO{\nu} \qquad (\nu>2)$, \label{+1_gener}
	
		where $\Ps{ijk}^{(-\nu)}$ can be expressed in terms of $\Om{ij}^{(-\nu)}$ using \Bi{8} (except when {$\nu=3,n$}). {For $\nu>3$, this case sets the boundary condition $\Ps{ijk}^{(3)}=0$, and for $\nu>n$ also $\Ps{ijk}^{(n)}=0$.} It includes both the case when $\bl$ is a {\em multiple} WAND (in the formal limit $\nu\to+\infty$) and asymptotically flat radiative spacetimes in higher dimensions (as we will discuss in the following, cf.~\cite{GodRea12}).
		
	\item $\Ps{ijk}=\OO{n} , \qquad \Om{ij}=\oo{n}$ , \label{+1_n}
	
	with $(n-3)\Ps{ijk}^{(n)}=2\Ps{[j}^{(n)}\delta_{k]i}$. {This case corresponds to the boundary condition $\Ps{ijk}^{(3)}=0$, $\Ps{ijk}^{(n)}\neq0$.}  It is compatible with the four-dimensional results of \cite{NP,Penrose63,Penrose65prs} (where $\nu=5$) for $n=4$.
	
	\item $\Ps{ijk}=\OO{3} , \qquad \Ps{i}=\oo{3} , \qquad \Om{ij}=\OO{\nu}$ \qquad $(n>4,\nu>3)$, \label{+1_3} 
	
		with $\Ps{i}=\OO{\nu}$ if $3<\nu\le 4$, and (using \eqref{nu>3}) $\Ps{i}=\OO{4}$ if $\nu>4$ {(in both cases the leading term of $\Ps{i}$ can be determined by the trace of \Bi{8}). This case corresponds to the boundary condition $\Ps{ijk}^{(3)}\neq0$.}  It is not permitted in 4D since $ \Ps{i}=0\Leftrightarrow\Ps{ijk}=0$ there \cite{OrtPraPra12rev} {and cannot be asymptotically flat, cf.~\cite{GodRea12}}.

\end{enumerate}

Only cases (\ref{+1_n}) and (\ref{+1_3}) are permitted if one assumes that asymptotically $\Ps{ijk}$ goes to zero more slowly than $\Om{ij}$.

Furthermore, from \Rc{e} we have
\be
 \ta{i}=\OO{1} , 
\ee
which with \eqref{comm_X} gives
\be
 {X^A=X^{A0}+\OO{1}} .
\label{X}
\ee

When the fall-off condition $\nu>3$ is assumed, thanks to \eqref{nu>3} we can strengthen the above results and those of section~\ref{subsec_commut} for the derivative operator as follows (assuming that each quantity has a power-like behaviour also at the subleading order): 
\beqn
	& & L_{1i}=\frac{l_{1i}}{r}+\OO{2}, \qquad L_{i1}=\frac{l_{i1}}{r}+\OO{2}, \qquad \M{i}{j}{k}=\frac{\m{i}{j}{k}}{r}+\OO{2} , \label{Ricci0_sublead} \\
	& & \xi^A_i=\frac{\xi^{A0}_i}{r}+\OO{2} , \qquad \om{i}=-l_{1i}+\OO{1}  \qquad\qquad (\nu>3) .
\eeqn
This will be useful in the following since many cases of interest have indeed $\nu>3$. {Note that using null rotations \eqref{nullrot} one can always choose a parallelly transported frame such that, e.g., $l_{1i}=0$ or $l_{i1}=0$. This may be convenient for particular computations but for the sake of generality we will keep our frame unspecified.}

{At this stage, knowing the $r$-dependence of the derivative operators at the leading order (eq.~\eqref{derivatives} with \eqref{xi}, \eqref{Rc0}, \eqref{X} and \eqref{U_R} or \eqref{U_R=0}) of course means also knowing the leading-order terms of the spacetime metric (however, to explicitly connect the metric and the Weyl tensor we would need to study higher-order terms).}
In the following, we will {analyze} in detail the above case (\ref{+1_gener}) (sections \ref{subsec_Ricci-1Weyl0}, \ref{subsubsec_R}, \ref{subsubsec_R=0}). For cases (\ref{+1_n}) and (\ref{+1_3}), we will only summarize the main results (sections~\ref{subsec_R_ii}, \ref{subsec_R_iii}, \ref{subsec_R=0_ii} and \ref{subsec_R=0_iii})  without giving intermediate steps since the method to obtain those is essentially the same as for case (\ref{+1_gener}).

\subsection{Ricci rotation coefficients of b.w. -1 and Weyl components of b.w. 0: {derivation for case (\ref{+1_gener})}}

\label{subsec_Ricci-1Weyl0}

The next step consists in the study of \Rc{a}, \Rc{j}, \Rc{m}, \Bi{5}, \Bi{12} and \eqref{comm_U}, also using the results of section~\ref{subsec_Rc0} above. It is convenient to start from \Rc{j} and \Bi{12} (since these do not contain $L_{11}$, $\M{i}{j}{1}$ and $U$). Let us first focus on \Rc{j} and consider the leading-order behaviour of the following quantities

\be
 \N{ij}=\OOp{\alpha} , \qquad \WD{ij}=\OOp{\beta} .
\label{def_alpha_beta}
\ee

By inspecting \Rc{j} we arrive at the following possibilities:
\begin{enumerate}

	\item\label{R} For $\tilde R\neq0$:
	\begin{enumerate}

		\item\label{R_>} $\alpha=1$, $\beta<0$, with $\N{ij}=-\frac{\tilde R}{2}\delta_{ij}r+\oop{}$ 
		
		\item\label{R_<} $\alpha<1$, $\beta=0$, with $\WD{ij}=-\tilde R\delta_{ij}+\oon$	
		
		\item\label{R_=} $\alpha\ge1$, $\beta=\alpha-1$.

	\end{enumerate}
	
	\item\label{R=0} For $\tilde R=0$:
	\begin{enumerate}

		\item\label{R=0_>} $\alpha=-1$, $\beta<-2$, with $\N{ij}=\OO{1}$
		
		\item\label{R=0=>} $\alpha\ge1$, $\beta=\alpha-1$
		
		\item\label{R=0=<} $\alpha<1$, $\alpha\neq-1$, $\beta=\alpha-1$.

	\end{enumerate}

\end{enumerate}

Let us also define the leading-order behaviour of 
\be
 \WD{ijkl}=\OOp{\beta_c} .
\ee
Now, in general, the leading-order term of eq.~\Bi{12} can be of order $\OOp{\beta_c-1}$, $\OOp{\beta-1}$, $\OOp{\alpha-\nu}$, $\OO{\nu-1}$, depending on the relative value of the parameters $\alpha$, $\beta_c$, $\beta$, $\nu$  {(recall that here we are restricting to case (i): $\Ps{ijk}=\OO{\nu}$, $\Om{ij}=\OO{\nu}$)}. It is easy to see that in the above cases (\ref{R_<}), (\ref{R_=}) and (\ref{R=0=>}) the leading term is either $\OOp{\beta_c-1}$ or $\OOp{\beta-1}$ (with possibly $\beta=\beta_c$). However, studying \Bi{12} at the leading order reveals that such cases (\ref{R_<}), (\ref{R_=}) and (\ref{R=0=>}) are in fact forbidden, since they all have $\beta\ge0$. Additionally, it shows that in case (\ref{R=0=<}) one has a stronger restriction $\alpha<-1$ (for $n=4$ eq.~\Bi{5} is also needed).
In the permitted cases, we can thus in general conclude 
\beqn
	& & \N{ij}=-\frac{\tilde R}{2}\delta_{ij}r+\oop{} \qquad \mbox{if } \tilde R\neq 0 , \label{N_R} \\
	& & \N{ij}=\OO{1} \qquad \mbox{if } \tilde R=0 . \label{N_R=0} 
\eeqn

Note also that in all the permitted cases we have have $\beta<0$. This enables us to use \Rc{a} to readily arrive at 
\beqn
	& & L_{11}=\tilde Rr+\oop{} \qquad \mbox{if } \tilde R\neq 0 , \label{L11_R} \\
	& & L_{11}={l_{11}+\oon} \qquad \mbox{if } \tilde R=0 , \label{L11_R=0} 
\eeqn
while \Rc{m} gives
\be
 \M{i}{j}{1}=\OOn ,
	\label{Mij1} 
\ee
and \eqref{comm_U} leads to
\beqn
	& & U=-\frac{\tilde R}{2}r^2+\oop{2} \qquad \mbox{if } \tilde R\neq 0 , \label{U_R} \\
	& & U={-l_{11}r+\oop{}} \qquad \mbox{if } \tilde R=0 . \label{U_R=0}
\eeqn

Thanks to the above discussion we can now study the consequences of \Bi{12}, as well as those of \Bi{5}, more systematically.  Clearly, from now on it will be necessary to distinguish case \ref{R}. ($\tilde R\neq0$) from case \ref{R=0}. ($\tilde R=0$).

\section{Case $\tilde R\neq0$}

\subsection{Case (i): $\Ps{ijk}=\OO{\nu}$, $\Om{ij}=\OO{\nu}$ ($\nu>2$)}

\label{subsubsec_R}

\subsubsection{Weyl components of b.w. 0}

\label{subsubsec_R_bw0}

At the leading order of \Bi{12}, we can have only (some of) the terms $\OOp{\beta_c-1}$/$\OOp{\beta-1}$, $\OOp{1-\nu}$. (From now on, it will be understood that $\WDS{ij}$ and $\WD{}$ have the same behaviour as $\WD{ijkl}$, {i.e., $\beta=\beta_c$,} except when stated otherwise.)

\begin{enumerate}[I.]

	\item If $1-\nu>\beta_c-1$ and $1-\nu>\beta-1$, eq.~\Bi{12} shows that necessarily $n=4$ and \Bi{5} then gives $\nu=5$. It also turns out that then $\beta_c=\beta=-4$, so that here we can thus have only
		
		\be
		{	\WD{ijkl}=\OO{4}, \qquad \WDA{ij}=\OO{4}, } \qquad \Om{ij}=\OO{5} \qquad (n=4) .
			\label{R_n=4_a}
		\ee

	\item In all remaining cases, at least one of the terms $\OOp{\beta_c-1}$, $\OOp{\beta-1}$ must appear at the leading order in \Bi{12}. Combing this with \Bi{5}, after some calculations and depending on the value of $\nu$ (and of $n$) one arrives at the following possible behaviours: 
		\begin{enumerate}
						
			\item{$\beta_c=-2$, $\nu=4$:} 
				\be
					\WD{ijkl}=\OO{2} , \qquad \WD{}=\oo{2}, \qquad \WDA{ij}=\oo{2} , \qquad \Om{ij}=\OO{4}  \qquad (n>4) . \label{R_i_bw0_a}
				\ee
				with $\Phi^{S(2)}_{ij}=\frac{\tilde R}{2}\Om{ij}^{(4)}$. Since in 4D $\WDS{ij}\propto\delta_{ij}$, this case is permitted only for $n>4$.

			\item{$\beta_c=-2$, $\nu>4$:} it follows from the last remark that here $\WDS{ij}$ becomes subleading. It turns out (by comparing \Bi{5} with the trace of \Bi{12}) that {the ranges $4<\nu<5$ and $4<\nu<6$ are} forbidden and we can identify three possible subcases, i.e.,
				
				\beqn
					& & \WD{ijkl}=\OO{2} , \quad \WDS{ij}=\oo{3}, \quad \WDA{ij}=\OO{3} , \quad \Om{ij}=\OO{5}  \qquad (n>{5}) , \label{R_i_bw0_b1} \\
					& & \WD{ijkl}=\OO{2} , \quad \WDS{ij}=\OOp{1-n}, \quad \WDA{ij}=\oop{1-n} , \quad \Om{ij}=\OO{n-1}  \qquad (n>{5}) , \label{R_i_bw0_b2} \\
					& & \WD{ijkl}=\OO{2} , \qquad \WDS{ij}=\OOp{2-\nu}, \qquad \WD{}=\oop{2-\nu}, \qquad \WDA{ij}=\oop{2-\nu} , \nonumber \\
					& & \qquad\qquad\qquad\qquad\qquad\qquad\qquad\qquad \Om{ij}=\OO{\nu}  \qquad (n>{5}, {\nu\ge6}, \nu\neq n+1) . \label{R_i_bw0_b3}
				\eeqn

		Here, $n>5$ since in 4D and 5D one has $\WD{ijkl}=0\Leftrightarrow\WDS{ij}=0$ \cite{PraPraOrt07}. In {\eqref{R_i_bw0_b1},} the ((anti)symmetric parts of the) trace of \Bi{12} (using \eqref{nu>3}) give $\WD{ijkl}^{(2)}b_{(jl)}=-\frac{\tilde R}{2}(n-4)\Om{ik}^{(5)}$ and $(n-4)\Phi_{ij}^{A(3)}=\WD{ikjl}^{(2)}b_{[kl]}$; {moreover, if $\nu>5$ then necessarily $\nu\ge6$. In \eqref{R_i_bw0_b2} and \eqref{R_i_bw0_b3}, we have instead} $\WD{ijkl}^{(2)}b_{(jl)}=0=\WD{ikjl}^{(2)}b_{[kl]}$. In \eqref{R_i_bw0_b2}, one finds $(2-n)\Phi^{S(n-1)}_{ij}+\Phi^{(n-1)}\delta_{ij}=\frac{\tilde R}{2}(n-4)\Om{ij}^{(n+1)}$, and $\Om{ij}$ can go to zero faster than indicated. In {\eqref{R_i_bw0_b3}} one has $(3-\nu)\Phi_{ij}^{S(\nu-2)}=\frac{\tilde R}{2}\Om{ij}^{(\nu)}(\nu-5)$ {(as obtained from \Bi{5})}.

		\item{$\beta_c=1-n$:} there is a difference between $n>4$ and $n=4$, i.e.,
				\beqn
					& & \mbox{if } n>4: \qquad  \WD{ijkl}=\OOp{1-n} , \qquad \WDA{ij}=\oop{1-n} , \qquad \Om{ij}=\OO{n-1} , \label{R_i_bw0_c} \\
					& & \mbox{if } n=4: \qquad \WD{ijkl}=\OO{3} , \qquad \WDA{ij}=\OO{3} , \qquad \Om{ij}=\OO{5} , \label{R_n=4_b}
				\eeqn
				
				with $\WD{ijkl}^{(3)}=2\Phi^{(3)}\delta_{j[k}\delta_{l]i}$ for $n=4$ and $(n-2)(n-3)\WD{ijkl}^{(n-1)}=4\Phi^{(n-1)}\delta_{j[k}\delta_{l]i}-2(n-3)\tilde R(\Om{j[k}^{(n+1)}\delta_{l]i}-\Om{i[k}^{(n+1)}\delta_{l]j})$ (which implies $(2-n)\Phi^{S(n-1)}_{ij}+\Phi^{(n-1)}\delta_{ij}=\frac{\tilde R}{2}(n-4)\Om{ij}^{(n+1)}$) for $n>4$. Note the different behaviour of the ``magnetic'' term $\WDA{ij}$. In both cases it is understood that $\Om{ij}$ can go to zero faster (or even vanish identically {-- for $n=4$ if $\nu>5$ then necessarily $\nu\ge6$}). In \eqref{R_n=4_b}, both $\WD{ijkl}$ and $\WDA{ij}$ can go to zero faster than indicated. The result of \eqref{R_n=4_a} can thus be understood as a subcase of \eqref{R_n=4_b} {-- for this reason \eqref{R_n=4_a} will not be considered anymore in the following}.

		\end{enumerate}

\end{enumerate}

We have not given explicitly the behaviour of $\Ps{ijk}$ in all the above cases since it always follows from point~(\ref{+1_gener}) of section~\ref{subsec_Rc0}. 
Note that not all values of $\nu$ are permitted. In particular, although we started from the weak assumption $\nu>2$, in the end we always have {either $\nu=4$ or $\nu\ge5$}.  Thanks to \eqref{nu>3} this enables us to specialize \eqref{N_R} to 
\be
	\N{ij}=-\frac{\tilde R}{2}\delta_{ij}r+\frac{\tilde R}{2}b_{ij}+\oon . \label{N_R_sublead}
\ee 	
Additionally, since in all permitted cases we have $\WD{}=\oo{2}$ and $\WDA{ij}=\oo{2}$   (or faster), eqs.~\eqref{L11_R}, \eqref{U_R} {and \eqref{Mij1}} can be specialized as 
\beqn
	& & L_{11}=\tilde Rr+l_{11}+\OO{1} , \label{L11_ref} \\
	& & U=-\frac{\tilde R}{2}r^2-l_{11}r+{\OOn} , \label{U_ref} \\
	& & \M{i}{j}{1}=\m{i}{j}{1}+\OO{1} \label{Mij1_ref} .   
\eeqn

{Using \eqref{N_R_sublead} in \Bi{5}, one is now able to refine all the ``$o$'' symbols in eqs.~\eqref{R_i_bw0_a}, \eqref{R_i_bw0_b2}, \eqref{R_i_bw0_b3} and \eqref{R_i_bw0_c} (but not in \eqref{R_i_bw0_b2}) by appropriate ``$O$'' symbols (e.g. $\WD{}=\oo{2}$ in \eqref{R_i_bw0_a} can be replaced by $\WD{}=\OO{3}$, etc). This will be taken into account explicitly in a summary in section~\ref{subsubsec_R_summary}.}

\subsubsection{Ricci rotation coefficients of b.w. -2 and Weyl components of b.w. -1}

\label{subsubsec_R_bw-1}

Let us analyze \Rc{f} and \Bi{6}, \Bi{9} and \Bi{1} in all the possible cases listed above, where we note that always $\nu\ge4$ (useful for the next comment). First, let us observe from \Bi{9} that if  $\Ps{ijk}'$ goes to zero more slowly than $\WD{ij}$ then necessarily it goes to zero as $\OO{2}$ (or faster). On the other hand, if $\Ps{ijk}'$ does {\em not} go to zero more slowly than $\WD{ij}$, we also conclude $\Ps{ijk}'=\OO{2}$ (or faster) since $\WD{ij}=\OO{2}$ (or faster) in all permitted cases. Thus, we always have $\Ps{ijk}'=\OO{2}$ (or faster), which enables one to use \Rc{f} (together with the second of \eqref{Ricci0_sublead} and \eqref{N_R_sublead})
to arrive at 
\be
	N_{i1}=\frac{\tilde R}{2}l_{i1}r+\OOn .
	\label{Ni1_R}
\ee

Thanks to this result we can now employ \Bi{6} together with \Bi{9} and arrive at the following results (where the various points are ``numbered'' so as to correspond to those of section~\ref{subsubsec_R_bw0}). From now on, it will be understood that $\Ps{i}'$ has the same behaviour as $\Ps{ijk}'$, except when stated otherwise.

\begin{enumerate}[(a)]

  \item $\Ps{ijk}'=\OO{2}$,
							
								with $\Ps{i}'^{(2)}=-\frac{\tilde R}{2}\Ps{i}^{(4)}$, and $\Ps{ijk}'^{(2)}$ can be expressed in terms of $\Om{ij}^{(4)}$ and $\WD{ijkl}^{(2)}$ using \Bi{6} (recall that $\Ps{ijk}^{(4)}$ and its trace $\Ps{i}^{(4)}$ can be expressed in terms of $\Om{ij}^{(4)}$, as observed in section~\ref{subsec_Rc0}).

	\item  For the three subcases we find, respectively,
			\beqn
					& & \Ps{ijk}'=\OO{2} , \qquad \Ps{i}'=\OO{3} , \\		
					& & \Ps{ijk}'=\OO{2} , \qquad \Ps{i}'=\OOp{1-n} , \\		
					& & \Ps{ijk}'=\OO{2}, \qquad \Ps{i}'=\OOp{2-\nu} ,		
			\eeqn
with $\Ps{ijk}'^{(2)}=-\WD{isjk}^{(2)}l_{s1}$, and where the behaviour of $\Ps{i}'$ has been obtained using \Bi{1}.

\item 
	
				\beqn
					& & \mbox{if } n>4: \qquad \Ps{ijk}'=\OOp{1-n} , \\		
					& & \mbox{if } n=4: \qquad \Ps{ijk}'=\OO{2} .
				\eeqn

\end{enumerate}

\subsubsection{Weyl components of b.w. -2}

\label{subsubsec_R_bw-2}

To conclude, let us study \Bi{4}. It will  be also useful to use \Bi{13}, for which the trace immediately tells us that the terms containing $\Om{ij}'$ cannot be leading over all the remaining terms in that equation (when $n>4$). Bearing this in mind, in the various cases listed above \Bi{4} leads to:

\begin{enumerate}[(a)]

	\item $\Om{ij}'=\OOn$,  					
			 
		with $\Om{ij}'^{(0)}=\left(\frac{\tilde R}{2}\right)^2\Om{ij}^{(4)}$. {(One can arrive at the same result also using \Bi{13}.)}

	\item\label{R_-2_c}  In the first case (eq.~\eqref{R_i_bw0_b1}), we find
			\be
					\Om{ij}'=\OO{1}  \qquad (\mbox{case \eqref{R_i_bw0_b1}}) ,
			\ee
	with $\Om{ij}'^{(1)}=-\left(\frac{\tilde R}{2}\right)^2\Om{ij}^{(5)}$, and for the second and third cases (eqs.~\eqref{R_i_bw0_b2}, \eqref{R_i_bw0_b3}) 
		\be
		 \Om{ij}'=\OO{2}  \qquad (\mbox{cases \eqref{R_i_bw0_b2}, \eqref{R_i_bw0_b3}}) .
		\ee
	The different behaviour in case~\eqref{R_i_bw0_b1} stems from \Bi{13} using the fact that $\WD{ijkl}^{(2)}b_{(jl)}\neq0$ when $\nu=5$. In case \eqref{R_i_bw0_b3} one has $\Om{ij}'^{(2)}=\WD{isjk}^{(2)}l_{s1}l_{k1}+\left(\frac{\tilde R}{2}\right)^2\Om{ij}^{(6)}$ {(recall that $\nu\ge6$, cf. section~\ref{subsubsec_R_bw0}).} For case \eqref{R_i_bw0_b2} one has simply $\Om{ij}'^{(2)}=\WD{isjk}^{(2)}l_{s1}l_{k1}$.

\item\label{R_-2_a} 
	
				\beqn
					& & \mbox{if } n>4: \qquad \Om{ij}'=\OOp{3-n} , \\		
					& & \mbox{if } n=4: \qquad \Om{ij}'=\OO{1} ,
				\eeqn
where $\Om{ij}'^{(n-3)}=\left(\frac{\tilde R}{2}\right)^2\Om{ij}^{(n+1)}$ for $n>4$. {(One can arrive at the same result also using \Bi{13}.)}

\end{enumerate}

It is clear that if $n>4$ and $\bl$ is a WAND (possible in cases (\ref{R_-2_a}) and (\ref{R_-2_c}) above) the fall-off of $\Om{ij}'$ will be faster since $\Om{ij}=0$ (in agreement with the results of \cite{OrtPraPra_prep} for multiple WANDs).

\subsubsection{Summary of case (i)}

\label{subsubsec_R_summary}

In all cases given here we have
\beqn
	& & \Om{ij}=\OO{\nu}  \qquad (\nu\ge4) , \nonumber \\
	& & \Ps{ijk}=\OO{\nu} .
\eeqn
These two equations will not be repeated every time below, where we will give only possible further restrictions on $\nu$. See also sections~\ref{subsubsec_R_bw0}--\ref{subsubsec_R_bw-2} for relations among the leading-order terms of various boost weight.

\begin{enumerate}[(a)]

	\item Here $n>4$ and
	
		\beqn
			& & \WD{ijkl}=\OO{2} , \qquad \WD{}=\OO{3}, \qquad \WDA{ij}=\OO{3}  \qquad (n>4, \, \nu=4) , \nonumber  \\
			& & \Ps{ijk}'=\OO{2} , \label{R_i_nu=4} \\
			& & \Om{ij}'=\OOn . \nonumber 
		\eeqn	
	
	The leading term at infinity is of order $r^0$ and it is of type N. At order $1/r^{2}$ the type becomes II(ad). This case does not seem of great physical interest since the frame components $\Om{ij}'$ do not decay near infinity. {In particular, it cannot describe asymptotically AdS spacetimes according to the definition of \cite{AshDas00} (this applies also to cases below and in sections \ref{subsec_R_ii} and \ref{subsec_R_iii} having $\WD{ijkl}=\OO{2}$ and/or $\Ps{ijk}=\OO{3}$).} Here $\bl$ cannot be a WAND.

	\item Here $n>5$ and we have three subcases. Generically {(case \eqref{R_i_bw0_b1})} we have
		
		\beqn
			& & \WD{ijkl}=\OO{2} , \qquad \WDS{ij}=\oo{3}, \qquad \WDA{ij}=\OO{3}  \qquad (n>5, \, {\nu=5 \mbox{ or } \nu\ge6}) ,  \nonumber  \\
			& & \Ps{ijk}'=\OO{2} , \qquad \Ps{i}'=\OO{3} , \label{R_i_nu>=5} \\
			& & \Om{ij}'=\OO{1} , \nonumber 
		\eeqn		
		where, however, if $\nu\ge6$ then $\WDS{ij}=\OO{4}$ and $\Om{ij}'=\OO{2}$. The leading term is thus of type N for {$\nu=5$} and of type type II(abd) for $\nu\ge6$. As a special subcase here $\bl$ can be a {multiple} WAND, cf. the results of \cite{OrtPraPra_prep}.

		If $\WD{ijkl}^{(2)}b_{[jl]}=0$ this becomes either

		\beqn
			& & \WD{ijkl}=\OO{2} , \qquad \WDS{ij}=\OOp{1-n}, \qquad \WDA{ij}=\OO{n}  \qquad (n>{5}, \, \nu\ge n+1) ,  \nonumber  \\
			& & \Ps{ijk}'=\OO{2} , \qquad \Ps{i}'=\OOp{1-n} , \\
			& & {\Om{ij}'=\OO{2}} , \nonumber 
		\eeqn	
		{which describes, in particular, the fall-off along a multiple WAND in Robinson-Trautman  Einstein spacetimes \cite{PodOrt06} (such as static Einstein black holes)} or (if {$6\le\nu<n+1$}, or $\nu>n+1$ but with $\Phi^{S(n-1)}_{ij}=0$) 
		
		\beqn
			& & \WD{ijkl}=\OO{2} , \quad \WDS{ij}=\OOp{2-\nu}, \quad \WD{}=\OOp{1-\nu}, \quad \WDA{ij}=\OOp{1-\nu}  \qquad (n>{5}, {\nu\ge 6}, \nu\neq n+1) , \nonumber  \\
			& & \Ps{ijk}'=\OO{2}, \qquad \Ps{i}'=\OOp{2-\nu} , \\
			& & {\Om{ij}'=\OO{2}} . \nonumber 
		\eeqn	
		
		{The leading term is of type II(abd) in both of the above two cases.}

	\item  This possibility arises when $\WD{ijkl}^{(2)}=0$ and includes the four-dimensional case. For $n>4$, we have
	
		\beqn
			& & \WD{ijkl}=\OOp{1-n} , \qquad \WDA{ij}=\OO{n}  \qquad (n>4, \, \nu\ge n+1) , \nonumber  \\
			& & \Ps{ijk}'=\OOp{1-n} , \label{R_i_betac=1-n} \\
			& & \Om{ij}'=\OOp{3-n} . \nonumber 
		\eeqn	
	
	The leading term at infinity is of order $1/r^{n-3}$ (provided $\Om{ij}^{(n+1)}\neq0$) and it is of type N. At order $1/r^{n-1}$ the type becomes {II(cd)} (II(bcd) if $\Om{ij}^{(n+1)}=0$). {In special cases $\bl$ can be a multiple WAND. This case thus includes the behaviour of algebraically special spacetimes along a non-degenerate geodesic multiple WAND {under the assumption $\WD{ijkl}^{(2)}=0$}, for which, however, $\Om{ij}'=\OOp{1-n}$ \cite{OrtPraPra_prep} (the $r$-dependence at the leading order has been worked out explicitly also for concrete examples such as Kerr-Schild-(A)dS geometries (with a non-degenerate Kerr-Schild vector) 
	\cite{MalPra11}, including rotating (A)dS black holes, and for Robinson-Trautman spacetimes with (A)dS asymptotics \cite{PodOrt06}, such as the Schwarzschild-Tangherlini (A)dS black hole).}

		For $n=4$, one has instead {(recall that \eqref{R_n=4_a} is a subcase of \eqref{R_n=4_b})}
		\beqn
			& & \WD{ijkl}=\OO{3} , \qquad \WDA{ij}=\OO{3}  \qquad (n=4, \, \nu\ge 5)  ,  \nonumber  \\
			& & \Ps{ijk}'=\OO{2} , \label{R_i_n=4} \\
			& & \Om{ij}'=\OO{1} . \nonumber 
		\eeqn
		{This is a special subcase of the standard 4D peeling \eqref{R_ii_n=4}.}

\end{enumerate}

\subsection{Case (ii): $\Ps{ijk}=\OO{n}$, $\Om{ij}=\oo{n}$}

\label{subsec_R_ii}

The behaviour of the Ricci rotation coefficients and derivative operators is the same as in case (i) and it will not be repeated here (in particular, \eqref{N_R}, \eqref{L11_R}, \eqref{Mij1}, \eqref{U_R} and \eqref{Ni1_R} still apply).

\subsubsection{Case $\beta_c=-2$, $n>5$}

All the following cases can occur only for $n>5$. In general, one has

\beqn
	& & \Om{ij}=\oo{n} , \nonumber \\	
	& & \Ps{ijk}=\OO{n} , \nonumber \\
	& & \WD{ijkl}=\OO{2}, \qquad \WDS{ij}=\OO{4} , \qquad \WDA{ij}=\OO{3} ,  \label{R_ii_betac=-2} \\
	& & \Ps{ijk}'=\OO{2} , \qquad \Ps{i}'=\OO{3}   ,\nonumber \\
	& & {\Om{ij}'=\OO{2}} , \nonumber 
\eeqn
with {$\WD{ijkl}^{(2)}b_{(jl)}=0$,} $(n-4)\Phi_{ij}^{A(3)}=\WD{ikjl}^{(2)}b_{[kl]}$ {and $\Ps{ijk}'^{(2)}=-\WD{isjk}^{(2)}l_{s1}$}. Here $\bl$ can be a single WAND, in special cases. For $\Ps{ijk}^{(n)}=0$ this reduces to \eqref{R_i_nu>=5} (with $\nu>n$).

If $\WD{ikjl}^{(2)}b_{[kl]}=0$ (but $\WD{ikjl}^{(2)}\neq0$) we have the subcase:

\beqn
	& & \Om{ij}=\OO{1-n} , \nonumber \\
	& & \Ps{ijk}=\OO{n} , \nonumber \\
	& & \WD{ijkl}=\OO{2}, \qquad \WDS{ij}=\OOp{1-n} , \qquad \WDA{ij}={\OO{n}} ,  \label{R_ii_2} \\
	& & \Ps{ijk}'=\OO{2} , \qquad \Ps{i}'=\OOp{2-n}    ,\nonumber \\
	& & {\Om{ij}'=\OO{2}} , \nonumber 
\eeqn
with $\Ps{i}'^{(n-2)}=\frac{\tilde R}{2}\Ps{i}^{(n)}$. {$\Om{ij}$ can go to zero faster than indicated.}

If, additionally, $\Phi_{ij}^{S(n-1)}=0$ we have, depending on the range of $\nu$, either

\beqn
	& & \Om{ij}=\OO{\nu} \qquad (n<\nu<2+n, \nu\neq n+1) , \nonumber \\
	& & \Ps{ijk}=\OO{n} , \nonumber \\
	& & \WD{ijkl}=\OO{2}, \qquad \WDS{ij}=\OOp{2-\nu} , \qquad \WD{}=\oop{2-\nu} , \qquad \WDA{ij}=\oop{2-\nu} , \label{R_ii_2_sub1} \\
	& & \Ps{ijk}'=\OO{2} , \qquad \Ps{i}'=\OOp{2-n}   ,\nonumber \\
	& &  {\Om{ij}'=\OO{2}} , \nonumber 
\eeqn
where the precise power of $r$ for both $\WD{}$ and $\WDA{ij}$ is given by $\max\{1-\nu,-n\}$, or
\beqn
	& & \Om{ij}=\OO{2-n} , \nonumber \\
	& & \Ps{ijk}=\OO{n} , \nonumber \\
	& & \WD{ijkl}=\OO{2}, \qquad \WDS{ij}=\OO{n} , \qquad \WDA{ij}=\OO{n} , \label{R_ii_2_sub2} \\
	& & \Ps{ijk}'=\OO{2} , \qquad \Ps{i}'=\OOp{2-n}   ,\nonumber \\
	& & {\Om{ij}'=\OO{2}} , \nonumber 
\eeqn
{where $\Om{ij}$ can go to zero faster than indicated.}

 In all the above cases the leading term is of type II(abd).

\subsubsection{Case $\beta_c<-2$, {$n>4$}}

If $\WD{ijkl}^{(2)}=0$ then \eqref{R_ii_2} reduces to

\beqn
	& & \Om{ij}=\OO{1-n} , \nonumber \\
	& & \Ps{ijk}=\OO{n} , \nonumber \\
	& & \WD{ijkl}=\OOp{1-n}, \qquad \qquad \WDA{ij}={\OO{n}} , \label{R_ii_betac<-2}  \\
	& & \Ps{ijk}'={\OOp{2-n}}    ,\nonumber \\
	& & {\Om{ij}'=\OOp{3-n}} , \nonumber 
\eeqn
	with $\Om{ij}'^{(n-3)}=\left(\frac{\tilde R}{2}\right)^2\Om{ij}^{(n+1)}$ and $(n-2)(n-3)\WD{ijkl}^{(n-1)}=4\Phi^{(n-1)}\delta_{j[k}\delta_{m]i}-2(n-3)\tilde R(\Om{j[k}^{(n+1)}\delta_{m]i}-\Om{i[k}^{(n+1)}\delta_{m]j})$. {The leading term is type N. If $\Ps{ijk}^{(n)}=0$ this reduces to \eqref{R_i_betac=1-n} with $\nu=n+1$.} Although the above fall-off looks very similar to the standard 4D peeling \eqref{R_ii_n=4}, an important difference for $n>4$ is that $\Om{ij}'^{(n-3)}\neq0$ {implies that} $\bl$ is not a WAND.

If $\Phi^{(n-1)}=0=\Om{ij}^{(n+1)}$ this becomes

\beqn
	& & \Om{ij}=\OO{2-n} , \nonumber \\
	& & \Ps{ijk}=\OO{n} , \nonumber \\
	& & \WD{ijkl}=\OO{n}, \qquad \WDA{ij}=\OO{n} ,  \\
	& & \Ps{ijk}'={\OOp{2-n}}    ,\nonumber \\
	& & {\Om{ij}'=\OOp{2-n}} . \nonumber 
\eeqn

Here the leading term is of type III. {$\Om{ij}$ can go to zero faster than indicated.}

In both the above cases we have $(n-3)\Ps{ijk}'^{(n-2)}=\tilde R\Ps{[j}^{(n)}\delta_{k]i}$.

\subsubsection{Case $n=4$}

{In four dimensions, we recover the standard asymptotic behaviour \cite{Penrose65prs,penrosebook2}, i.e.,}
\beqn
	& & \Om{ij}=\OO{\nu} \qquad (\nu\ge 5) , \nonumber \\
	& & \Ps{ijk}=\OO{4} , \nonumber \\
	& & \WD{ijkl}=\OO{3}, \qquad \WDA{ij}=\OO{3},  \label{R_ii_n=4} \\
	& & \Ps{ijk}'={\OO{2}}   ,\nonumber \\
	& & \Om{ij}'={\OO{1}} . \nonumber 
\eeqn

{In our study, the condition $\nu\ge 5$ followed by analyzing the Ricci and Bianchi equations (where we initially only assumed $\nu>2$), thanks to $\tilde R\neq0$. Additionally, we observe that} if $\nu>5$ then necessarily $\nu\ge6$. {For $\Ps{ijk}^{(4)}=0$ this case reduces to \eqref{R_i_n=4}.}

\subsection{Case (iii): $\Ps{ijk}=\OO{3}$, $\Om{ij}=\oo{3}$ ($n>4$)}

\label{subsec_R_iii}

Again the behaviour of the Ricci rotation coefficients and derivative operators is the same as in case (i).\footnote{To arrive at \eqref{Ni1_R} in the present case one needs to use also \eqref{nu>3} and \eqref{N_R_sublead}, and thus to observe that although \Bi{9} gives $\Ps{ijk}'=\OO{1}$, from its trace one gets $\Ps{i}'=\OO{2}$ (see also \eqref{R_iii_1}--\eqref{R_iii_3} below).}

\subsubsection{{Case $\beta_c=-2$}}

Here in general one has {($n\ge5$)}

\beqn
	& & \Om{ij}=\OO{4} , \nonumber \\
	& & \Ps{ijk}=\OO{3} , \qquad \Ps{i}=\OO{4} , \nonumber \\
	& & \WD{ijkl}=\OO{2}, \qquad \WD{}={\OO{3}} , \qquad \WDA{ij}={\OO{3}} ,  \label{R_iii_1} \\
	& & \Ps{ijk}'=\OO{1} , \qquad  \Ps{i}'=\OO{2} ,\nonumber \\
	& & {\Om{ij}'=\OOn} , \nonumber 
\eeqn
with $\Phi^{S(2)}_{ij}=\frac{\tilde R}{2}\Om{ij}^{(4)}$,  $\Ps{ijk}'^{(1)}=\frac{\tilde R}{2}\Ps{ijk}^{(3)}$, {$\Ps{i}'^{(2)}$ can be expressed in terms of $\Om{ij}^{(4)}$ and $\Ps{ijk}^{(3)}$ thanks to \Bi{6},} $\Om{ij}'^{(0)}=\left(\frac{\tilde R}{2}\right)^2\Om{ij}^{(4)}$ and
\be
 (n-4)\Phi_{ki}^{A(3)}=\WD{klij}^{(2)}b_{[lj]}+\xi^{A0}_j\Ps{[ki]j,A}^{(3)}+2l_{1j}\Ps{[ki]j}^{(3)}+\Ps{[ki]l}^{(3)}\m{l}{j}{j}+\Ps{jl[k}^{(3)}\m{l}{i]}{j}+\tilde R\Om{j[k}^{(4)}b_{i]j} . \nonumber 
\ee
The leading term is type N. In the limit $\Ps{ijk}^{(3)}=0$ this reduces to case \eqref{R_i_nu=4}.

If $\Om{ij}$ has a faster fall-off one finds for $n>5$ {(as in section~\ref{subsubsec_R} the range $4<\nu<5$ is forbidden by imposing \Bi{5} and \Bi{12}; {see section~\ref{subsubsec_R_iii_n=5} for the case $n=5$})}

\beqn
	& & \Om{ij}=\OO{\nu} \qquad {(\nu\ge5)} , \nonumber \\
	& & \Ps{ijk}=\OO{3} , \qquad \Ps{i}=\OO{4} , \nonumber \\
	& & \WD{ijkl}=\OO{2}, \qquad \WDS{ij}=\OO{3} , \qquad \WD{}={\OO{4}} ,\qquad \WDA{ij}=\OO{3} , \qquad\qquad (n>5)  \label{R_iii_2} \\
	& & \Ps{ijk}'=\OO{1} , \qquad  \Ps{i}'=\OO{2} ,\nonumber \\
	& & {\Om{ij}'=\OO{1}} , \nonumber 
\eeqn
where {from \Bi{5}} $\Phi_{ij}^{S(3)}=-\Ps{(ij)l}^{(3)}l_{l1}$, {from \Bi{12}} 
\beqn
 & & (n-4)\Phi_{ki}^{A(3)}=\WD{klij}^{(2)}b_{[lj]}+\xi^{A0}_j\Ps{[ki]j,A}^{(3)}+2l_{1j}\Ps{[ki]j}^{(3)}+\Ps{[ki]l}^{(3)}\m{l}{j}{j}+\Ps{jl[k}^{(3)}\m{l}{i]}{j} , \nonumber \\
 & &  {}-\WD{klij}^{(2)}b_{(lj)}=\xi^{A0}_j\Ps{(ki)j,A}^{(3)}+[(n-6)l_{j1}+2l_{1j}]\Ps{(ki)j}^{(3)}+\Ps{(ki)l}^{(3)}\m{l}{j}{j}+(2\Ps{l(k|j}^{(3)}+\Ps{jl(k|}^{(3)})\m{l}{|i)}{j}+\frac{\tilde R}{2}(n-4)\Om{ik}^{(5)} , \nonumber 
\eeqn
and $\Om{ij}'^{(1)}$ can be expressed (using the trace of \Bi{13}) in terms of $\Om{ik}^{(5)}$ and $\Ps{ijk}^{(3)}$.

The leading term is of type III(a) and $\bl$ can be a single WAND. If $\Ps{ijk}^{(3)}=0$ this reduces to \eqref{R_i_nu>=5} for $5\le\nu\le n$ and to \eqref{R_ii_betac=-2} for $\nu>n$.

\subsubsection{{Case $\beta_c<-2$}}

\label{subsubsec_R_iii_n=5}

{For $n=5$, or for $n>5$ with $\WD{ijkl}^{(2)}=0$, instead of \eqref{R_iii_2} one has}

\beqn
	& & \Om{ij}=\OO{\nu} \qquad (\nu\ge5) , \nonumber \\
	& & \Ps{ijk}=\OO{3} , \qquad {\Ps{i}=\OO{4} ,} \nonumber \\
	& & \WD{ijkl}=\OO{3}, \qquad \WD{}={\OO{3}} , \qquad \WDA{ij}=\OO{3} , \qquad\qquad (n{\ge5})  \label{R_iii_3} \\
  & & \Ps{ijk}'=\OO{1} , \qquad  \Ps{i}'=\OO{2} ,\nonumber \\
	& & {\Om{ij}'=\OO{1}}  , \nonumber 
\eeqn
where $\WD{ijkl}^{(3)}$ can be expressed in terms of $\Om{ij}^{(5)}$ and $\Ps{ijk}^{(3)}$ using \Bi{12} (or \Bi{13}). The leading term is of type III(a). Again $\Ps{ijk}'^{(1)}=\frac{\tilde R}{2}\Ps{ijk}^{(3)}$.

{All of the above results for the case $\tilde R\neq0$ are summarized in table~\ref{tab_R}}.

\begin{footnotesize}
\begin{sidewaystable}[!htbp]
\footnotesize
 \begin{center}
   \begin{tabular}{|c ccccccccc|}
    \hline case & $\Omega_{ij}$ & $\Psi_{ijk}$, $\Psi_i$ & $\Phi_{ijkl}$ & $\Phi^S_{ij}$, $\Phi$ & $\Phi^A_{ij}$& $\Psi'_{ijk}$, $\Psi'_i$ & $\Omega'_{ij}$ & restrictions & comments \\ \hline

 (i) (a) &  $r^{-4}$ & {$r^{-4}$} & $r^{-2}$& $r^{-2}$, $r^{-3}$&  $r^{-3}$ & $r^{-2}$ &  ${\OOn}$ & 
$\nu=4$& $\bl$ not a WAND \\ [1mm]\hline
(i) (b) &  {$r^{-5}$}  & {{$r^{-5}$}} &  $r^{-2}$ & ${ o}(r^{-3})$ & $r^{-3}$ & $r^{-2}$,  $r^{-3}$  & $r^{-1}$ &  $n>5$, {$\nu= 5$} & {$\bl$ not a WAND} \\[1mm]
 &  $r^{-\nu}$  & {$r^{-\nu}$} &  $r^{-2}$ & $r^{-4}$ & $r^{-3}$ & $r^{-2}$,  $r^{-3}$  & $r^{-2}$ &  $n>5$, $\nu\geq 6$ &  \\[1mm]
 & {$r^{-\nu}$}  & {$r^{-\nu}$} & $r^{-2}$  &  $r^{1-n}$ &  $r^{-n}$ & $r^{-2}$,  $r^{1-n}$  & $r^{-2}$ & $n>5$, $\nu\geq n+1$& includes RT\\[1mm]
& $r^{-\nu}$ & {$r^{-\nu}$} & $r^{-2}$&  $r^{2-\nu}$,  $r^{1-\nu}$  & $r^{1-\nu}$ & $r^{-2}$,  $r^{2-\nu}$  & $r^{-2}$ & $n>5$, $\nu\geq 6$, $\nu\not= n+1$ & 

 \\ [1mm]\hline
(i) (c) 
& $r^{-n-1}$  & {$r^{-n-1}$} &  $r^{1-n}$  & & $r^{-n}$ & $r^{1-n}$ & $r^{3-n}$ & 
{$\nu= n+1$}& {$\bl$ not a WAND} \\ [1mm] 

& $r^{-\nu}$  & $r^{-\nu}$&  $r^{1-n}$  & & $r^{-n}$ & $r^{1-n}$ & $o(r^{3-n})$ & $\nu> n+1$& includes KS (A)dS \\ [1mm] 
 & $r^{-\nu}$ & {$r^{-\nu}$} &$r^{-3}$ & & $r^{-3}$&$r^{-2}$ &$r^{-1}$ &$n=4$, $\nu\geq 5$ & \\[1mm]
\hline \hline


(ii) & {${o}(r^{-n})$} & {$r^{-n}$} &$r^{-2}$ & $r^{-4}$  & $r^{-3}$   &$r^{-2}$, $r^{-3}$     &$r^{-2}$    & {$n>5$} &  \\[1mm]
& $r^{-n-1}$  & {$r^{-n}$} & $r^{-2}$  &${r^{1-n}}$   & $r^{-n}$   &$r^{-2}$  , $r^{{2-n}}$   &$r^{-2}$   & $n>5$, {$\nu\ge n+1$} & \\[1mm]
& $r^{-\nu}$   & {$r^{-n}$} & $r^{-2}$  &$r^{2-\nu}$, {${o}(r^{2-\nu})$}     &{${o}(r^{2-\nu})$}   & $r^{-2}$, $r^{{2-n}}$    &$r^{-2}$   &${5<}n<\nu<n+2$, $\nu\not=n+1$ & {$\bl$ not a WAND} \\[1mm]
&$r^{-n-2}$   & {$r^{-n}$} &$r^{-2}$   &$r^{-n}$   & $r^{-n}$  &$r^{-2}$, $r^{{2-n}}$     &$r^{-2}$   & $n>5$, {$\nu\ge n+2$} &\\[1mm]\hline
& $r^{-n-1}$  & {$r^{-n}$} & ${r^{1-n}}$  &${r^{1-n}}$   & $r^{-n}$   &$r^{{2-n}}$    &$r^{3-n}$   & {$\nu=n+1$} & $\bl$ not a WAND \\[1mm]
& $r^{-n-2}$  & {$r^{-n}$} & $r^{-n}$  &$r^{-n}$   & $r^{-n}$   &$r^{{2-n}}$    &$r^{{2-n}}$   &{$\nu\ge n+2$} & \\[1mm]
 & $r^{-\nu}$  & $r^{-4}$ & $r^{-3}$  &$r^{-3}$   & $r^{-3}$   &$r^{-2}$    &$r^{-1}$   & $n=4$, $\nu\geq 5$& \\[1mm]\hline\hline
%
%


(iii) & $r^{-4}$  & {$r^{-3}$,  $r^{-4}$} & $r^{-2}$  &$r^{-2}$, $r^{-3}$    & $r^{-3}$   &$r^{-1}$, $r^{-2}$  &${\OOn}$   & 
 {$\nu=4$} & {$\bl$ not a WAND} \\[1mm]
 & $r^{-\nu}$  & {$r^{-3}$,  $r^{-4}$} & $r^{-2}$  &$r^{-3}$, $r^{-4}$    & $r^{-3}$   &$r^{-1}$, $r^{-2}$     &$r^{-1}$   & $n>5$, $\nu\geq 5$ & \\[1mm]
 & $r^{-\nu}$  & {$r^{-3}$,  $r^{-4}$} & $r^{-3}$  &$r^{-3}$   & $r^{-3}$   &$r^{-1}$, $r^{-2}$     &$r^{-1}$   & $n\geq5$, $\nu\geq 5$ & \\[1mm]\hline\hline

\end{tabular}
  \caption{Fall-off behaviour of the Weyl tensor in the presence of a cosmological constant ({$\tilde R\not=0$}). We list here in a compact way the cases summarized in sections \ref{subsubsec_R_summary}, \ref{subsec_R_ii}, \ref{subsec_R_iii}.
	{Recall that} the cases (i), (ii) and (iii) differ by the fall-off of the component $\Psi_{ijk}$. Whenever there is just one power of $r$ in the column for $\Phi^S_{ij}$
	and $\Phi$ {(the 5th column)}, it means that these two quantities have the same fall-off (the same holds for $\Psi_{ijk}$, $\Psi_i$ and $\Psi'_{ijk}$, $\Psi'_i$ -- the 3rd and the 7th column, respectively), while when the column is empty it means that both $\Phi^S_{ij}$ and $\Phi$ have same fall-off as $\Phi_{ijkl}$. It is always understood that $n>4$ except when we explicitly indicate $n=4$ (last but one column). The shortcuts RT and KS stands for Robinson-Trautman and Kerr-Schild spacetimes, respectively (last column).
\label{tab_R}}
 \end{center}
\end{sidewaystable}
\end{footnotesize}

\section{Case $\tilde R=0$}

\label{sec_R=0}

\subsection{Case (i): $\Ps{ijk}=\OO{\nu}$, $\Om{ij}=\OO{\nu}$ ($\nu>2$)}

\label{subsubsec_R=0}

\subsubsection{Weyl components of b.w. 0}

\label{subsubsec_R=0_bw0}

In this case, at the leading order of \Bi{12} we can have only (some of) the terms $\OOp{\beta_c-1}$, $\OOp{\beta-1}$, $\OOp{-1-\nu}$. The same is true for the antisymmetric part of \Bi{5}, while the leading-order terms of the symmetric part of \Bi{5} can only be $\OOp{\beta_c-1}$, $\OOp{\beta-1}$, $\OOp{-\nu}$. Here, we are mainly interested in studying the case when the leading terms of \Bi{12} are $\OOp{\beta_c-1}$, $\OOp{\beta-1}$, i.e., $\beta_c>-\nu$ or $\beta>-\nu$. (In all the remaining cases, the asymptotic behaviour of b.w. zero components can be represented by $\WD{ijkl}=\OO{\nu}$, $\WDA{ij}=\OO{\nu}$, $\Om{ij}=\OO{\nu}$, with  $\nu>2$. The behaviour of higher b.w. components is given in section~\ref{subsubsec_R=0_i_special} below.)

By combining \Bi{12} and \Bi{5} we arrive at the following possibilities, also depending on the value of $\nu$ and of $n$:

\begin{enumerate}[(A)]\label{1st}

\item{$\beta_c=-2$:}\label{R=0_-2} there are several possibilities, i.e.,
	\begin{enumerate}[{A}1:]
					\item 
					\be
						\WD{ijkl}=\OO{2} , \qquad \WDS{ij}=\oo{2}, \qquad \WDA{ij}=\oo{2} , \qquad \Om{ij}=\OO{\nu}  \qquad (n>{5}, \, 2<\nu\le3) . \label{R=0_nu<=3}
					\ee

				\item 
					\beqn
						& & \WD{ijkl}=\OO{2} , \qquad \WDS{ij}=\OO{3} , \qquad \WD{}=\OO{\nu}, \qquad \WDA{ij}=\OO{3} , \qquad \Om{ij}=\OO{\nu} \nonumber \\
						& & \hspace{11cm} \qquad (n>{5}, \, 3<\nu<4) .
					\eeqn
						The (anti)symmetric parts of the trace of \Bi{12} (using \eqref{nu>3}) give $(n-4)\Phi_{ij}^{A(3)}=\WD{ikjl}^{(2)}b_{[kl]}$ {and $(n-6)\Phi_{ki}^{S(3)}=\WD{klij}^{(2)}b_{(lj)}$.}
			In the special case $\WD{ikjl}^{(2)}b_{[kl]}=0$ thus $\WDA{ij}$ goes to zero faster, namely $\WDA{ij}=\OO{\nu}$.

		 \item 
					\beqn
						& & \WD{ijkl}=\OO{2} , \qquad \WDS{ij}=\OO{3}, \qquad \WD{}=\OO{4} , \qquad \WDA{ij}=\OO{3} , \qquad \Om{ij}=\OO{4} \nonumber \\
						& & \hspace{11cm} (n>5) . \label{R=0_B_nu=4}
					\eeqn			
					As above $(n-4)\Phi_{ij}^{A(3)}=\WD{ikjl}^{(2)}b_{[kl]}$ and $(n-6)\Phi_{ki}^{S(3)}=\WD{klij}^{(2)}b_{(lj)}$ but here with the latter \Bi{5} further gives {$\WD{klij}^{(2)}b_{(lj)}=-(n-6)(l_{11}\Om{ki}^{(4)}+\frac{1}{2}X^{A0}\Om{ki,A}^{(4)}+\Om{s(k}^{(4)}\m{s}{i)}{1})$.} 
								Here $\Om{ij}$ can go to zero faster than indicated, i.e., $\Om{ij}=\OO{\nu}$ with $\nu>4$, but in that case {clearly} also $\WDS{ij}$ does (namely, $\WDS{ij}=\OOp{1-\nu}$ for $4<\nu<5$ and $\WDS{ij}=\OO{4}$ for $\nu\ge5$ -- in particular, for $\nu>5$ the symmetric part of \Bi{5} gives $\Phi_{ij}^{S(4)}$ in terms of $\Phi_{ij}^{A(3)}$).

					If $\WD{ikjl}^{(2)}b_{[kl]}=0$ we obtain the following two subcases, depending on whether $\nu\neq n$ or $\nu=n$.

				 \item 
					\beqn
						& & \WD{ijkl}=\OO{2} , \qquad \WDS{ij}=\OOp{1-\nu}, \qquad \WD{}=\OO{\nu} , \qquad \WDA{ij}=\OO{\nu} , \qquad \Om{ij}=\OO{\nu} \nonumber \\
						& & \hspace{11cm} (n>{5}, \, \nu\ge4, \, \nu\neq n),
					\eeqn			
					{with $\WD{ikjl}^{(2)}b_{[kl]}=0$ and $(n-6)\Phi_{ki}^{S(3)}=\WD{klij}^{(2)}b_{(lj)}$ (if $\nu=4$) or $\WD{klij}^{(2)}b_{(lj)}=0$ (if $\nu>4$).} For $\nu>n$ this can be seen as a subcase of \eqref{R=0_B_beta=1-n} with $\Phi^{(n-1)}=0$.

		\item 
					\be
						\WD{ijkl}=\OO{2} , \qquad \WDS{ij}=\OOp{1-n}, \qquad \WDA{ij}=\OO{n} , \qquad \Om{ij}=\OO{n}  \qquad (n>5) , \label{R=0_B_beta=1-n}
					\ee				
					 {with $\WD{ikjl}^{(2)}b_{[kl]}=0$ and $\WD{klij}^{(2)}b_{(lj)}=0$.} $\Om{ij}$ can go to zero faster than indicated, with no effect on {the fall-off of $\WDS{ij}$. If $\nu>n$ then $(2-n)\Phi^{S(n-1)}_{ij}+\Phi^{(n-1)}\delta_{ij}=0$.}

  \end{enumerate}

\item{$\beta_c=-n/2$:}\label{R=0_-n/2} 
				\be
					\WD{ijkl}=\OO{n/2} , \qquad \WD{}=\OO{\nu} , \qquad \WDA{ij}=\OO{\nu} , \qquad \Om{ij}=\OO{\nu}  \qquad (n>4, \, \frac{n}{2}<\nu\le1+\frac{n}{2}) ,
				\ee					
		with $(n-4)\WD{ijkl}^{(n/2)}=4(\Phi^{S(n/2)}_{i[l}\delta_{k]j}-\Phi^{S(n/2)}_{j[l}\delta_{k]i})$. 	Note that here $\Om{ij}$ cannot become $\oo{n/2-1}$ as long as $\WD{ijkl}=\OO{n/2}$. In the special case $\nu=1+n/2$, from \Bi{5} we obtain $(n-2)\Phi_{ij}^{S(n/2)}=-2X^{A0}\Om{ij,A}^{(n/2+1)}-(n-2)l_{11}\Om{ij}^{(n/2+1)}-4\Om{s(j}^{(n/2+1)}\m{s}{i)}{1}$, {while for $\frac{n}{2}<\nu<1+\frac{n}{2}$ we have $X^{A0}\Om{ij,A}^{(\nu)}+(\nu-2)l_{11}\Om{ij}^{(\nu)}+2\Om{s(j}^{(\nu)}\m{s}{i)}{1}=0$.}

\item{$\beta_c=1-n$:}\label{R=0_1-n} similarly as in section~\ref{subsubsec_R}, one has to distinguish between the cases $n>4$ and $n=4$, i.e.,
				\beqn
					& & \mbox{if } n>4: \qquad  \WD{ijkl}=\OOp{1-n} , \qquad \WDA{ij}=\oop{1-n} , \qquad \Om{ij}=\OO{\nu} \qquad (\nu>n-1) , \label{R=0_A}  \\
					& & \mbox{if } n=4: \qquad \WD{ijkl}=\OO{3} , \qquad \WDA{ij}=\OO{3} , \qquad \Om{ij}=\OO{\nu} \qquad (\nu>3) , \label{R=0_n=4}
				\eeqn
				{with (for $n\ge4$) $(n-2)(n-3)\WD{ijkl}^{(n-1)}=4\Phi^{(n-1)}\delta_{j[k}\delta_{l]i}$ and $(2-n)\Phi^{S(n-1)}_{ij}+\Phi^{(n-1)}\delta_{ij}=0$.} In~\eqref{R=0_A} we have $\WDA{ij}=\OO{\nu}$ for $n-1<\nu<n$ and $\WDA{ij}=\OO{n}$ for $\nu\ge n$.

\end{enumerate}

Again, see point~(\ref{+1_gener}) of section~\ref{subsec_Rc0} for the behaviour of $\Ps{ijk}$ in all the above cases. As shown above, in all cases except \eqref{R=0_nu<=3} we have $\nu>3$, which enables us (thanks to \eqref{nu>3}) to specialize \eqref{N_R=0} to 
\be
	\N{ij}=\frac{n_{ij}}{r}+\OO{2} \qquad \mbox{(except for \eqref{R=0_nu<=3})}.  \label{Nij_ref_R=0}  
\ee

Similarly as for $\tilde R\neq0$ (cf. section~\ref{subsubsec_R_bw0}), since in all permitted cases one has $\WD{}=\oo{2}$ and $\WDA{ij}=\oo{2}$,   for $L_{11}$, $U$, $\M{i}{j}{1}$ one obtains the refined equations that follow by setting $\tilde R=0$ in \eqref{L11_ref}, \eqref{U_ref}, \eqref{Mij1_ref} (in contrast to \eqref{Nij_ref_R=0} this applies also when $2<\nu\le3$).

\subsubsection{Ricci rotation coefficients of b.w. -2 and Weyl components of b.w. -1}

\label{subsubsec_R=0_bw-1}
	
Let us analyze \Rc{f} and \Bi{6}, \Bi{9} and \Bi{1} in all the possible cases listed above. Similarly as in section~\ref{subsubsec_R_bw-1}, it is easy to conclude from \Bi{9} that we always have $\Ps{ijk}'=\OO{2}$ (or faster, see more details below), which enables one to use \Rc{f} to obtain
\be
	N_{i1}=\OOn .
	\label{R=0_Ni1}
\ee
Using \Bi{9}, \Bi{6}, \Bi{1} one arrives at the following results (the numbering corresponds to that of section~\ref{subsubsec_R=0_bw0}).

\begin{enumerate}[(A)]

  \item For the five subcases we find, respectively,
		\begin{enumerate}[{A}1:] 
		
			\item $\Ps{ijk}'=\OO{2}$.  					
			
			\item $\Ps{ijk}'=\OO{2}$, $\Ps{i}'=\OO{3}$.
				
			\item $\Ps{ijk}'=\OO{2}$, $\Ps{i}'=\OO{3}$.  									
			
			\item $\Ps{ijk}'=\OO{2}$, $\Ps{i}'=\OOp{1-\nu}$.  					
				
			\item $\Ps{ijk}'=\OO{2}$, $\Ps{i}'=\OOp{1-n}$.  									
			
		\end{enumerate}
	
In all cases {except A1} we have $\Ps{ijk}'^{(2)}=-\WD{isjk}^{(2)}l_{s1}$ {(in case A1, if $\nu=3$ then \Bi{6} gives $\Ps{ijk}'^{(2)}$ in terms of $\Om{ij}^{(3)}$, $\Ps{ijk}^{(3)}$, and $\WD{isjk}^{(2)}$).}

  \item We have $\Ps{ijk}'=\OO{n/2}$ for any $n\ge6$, and for $n=5$ provided $3<\nu\le\frac{7}{2}$ (in both cases \Bi{9} enables one to express $\Ps{ijk}'^{(n/2)}$ in terms of {$\Phi_{ij}^{S(n/2)}$}). If, instead, $n=5$ and $\frac{5}{2}<\nu\le3$ we have $\Ps{ijk}'=\OO{2}$.

	\item 
	
				\beqn
					& & \mbox{if } n>4: \qquad \Ps{ijk}'=\OOp{1-n} , \\		
					& & \mbox{if } n=4: \qquad \Ps{ijk}'=\OO{2} .
				\eeqn
				
				For $n>4$, \Bi{9} {gives $(n-3)\Ps{ijk}'^{(n-1)}=2\Ps{[j}'^{(n-1)}\delta_{k]i}$, with $(n-2)\Ps{i}'^{(n-1)}=-(n-1)\WD{}^{(n-1)}l_{1i}-\xi^{A0}_i\WD{,A}^{(n-1)}$.}

\end{enumerate}

\subsubsection{Weyl components of b.w. -2}

\label{subsubsec_R=0_bw-2}

Using  \Bi{4} and \Bi{13}  we arrive at

\begin{enumerate}[(A)]

\item For the five subcases we find, respectively,
		\begin{enumerate}[{A}1:] 
		
			\item $\Om{ij}'=\OOp{\sigma}$, {with $-2\le\sigma<-1$ (the precise value of $\sigma$ depends on the values taken by $\nu$ and $\beta$ -- recall \eqref{def_alpha_beta}).}


				\item[{A}2--{A}5:]  $\Om{ij}'=\OO{2}$, with $\Om{ij}'^{(2)}=-3l_{11}\Phi_{ij}^{S(3)}-X^{A0}\Phi_{ij,A}^{S(3)}-2\Phi_{s(j}^{S(3)}\m{s}{i)}{1}-\Ps{(ij)k}'^{(2)}l_{k1}$ (note that in some of these cases $\Phi_{ij}^{S(3)}=0$).	
	
	\end{enumerate}

	  \item In all cases ($n\ge5$) we have
			\be
			  \Om{ij}'=\OOp{1-n/2} ,
			\ee 
		with $(n-4)\Om{ij}'^{(n/2-1)}=-nl_{11}\Phi_{ij}^{S(n/2)}-2X^{A0}\Phi_{ij,A}^{S(n/2)}-4\Phi_{s(j}^{S(n/2)}\m{s}{i)}{1}$. In the special case $\nu=1+n/2$ this can be written in terms of $\Om{ij}^{(n/2+1)}$ using the form of $\Phi_{ij}^{S(n/2)}$ given in the above section \ref{subsubsec_R=0_bw0}.

	\item 
	
				\beqn
					& & \mbox{if } n>4: \qquad \Om{ij}'=\oop{2-n} , \\		
					& & \mbox{if } n=4: \qquad \Om{ij}'=\OO{1}.
				\eeqn
				
				To obtain the above behaviour, in the $n>4$ case it is also necessary to recall that at the leading order $\WDS{ij}\propto\delta_{ij}$ (cf. section~\ref{subsubsec_R=0_bw0}).

\end{enumerate}

\subsubsection{Summary of case (i)}

\label{subsubsec_R=0_summary}

In all cases given here we have
\beqn
	& & \Om{ij}=\OO{\nu}  \qquad (\nu>2) , \nonumber \\
	& & \Ps{ijk}=\OO{\nu} .
\eeqn
This will not be repeated every time below, where we will give only possible further restrictions on $\nu$. See also sections~\ref{subsubsec_R=0_bw0}--\ref{subsubsec_R=0_bw-2} for relations among the leading-order terms of various boost weight.

\begin{enumerate}[(A)]

	\item Here we have $n>5$ and the following possible behaviours ({cf.~section~\ref{subsubsec_R=0_bw0} for a few further special subcases)}.
	
	\begin{enumerate}[{A}1:] 
		
	\item 
		
		\beqn
			& & \WD{ijkl}=\OO{2} , \qquad \WDS{ij}=\oo{2}, \qquad \WDA{ij}=\oo{2}  \qquad (n>{5}, \, 2<\nu\le3) , \nonumber  \\
			& & \Ps{ijk}'=\OO{2} , \\
			& & \Om{ij}'= \OOp{\sigma} \qquad {(-2\le\sigma<-1)} . \nonumber 
		\eeqn

	\item 
		
		\beqn
			& & \WD{ijkl}=\OO{2} , \quad \WDS{ij}=\OO{3} , \quad \WD{}=\OO{\nu}, \quad \WDA{ij}=\OO{3}  \quad (n>5, \, 3<\nu<4) , \nonumber  \\
			& & \Ps{ijk}'=\OO{2} , \quad \Ps{i}'=\OO{3} , \label{R=0_i_3<nu<4} \\
			& & \Om{ij}'=\OO{2} . \nonumber 
		\eeqn

	\item 
				
		\beqn
			& & \WD{ijkl}=\OO{2} , \qquad \WDS{ij}=\OO{3} , \qquad \WD{}=\OO{4} , \qquad \WDA{ij}=\OO{3}  \qquad (n>5, \, \nu{\ge 4}) , \nonumber  \\
			& & \Ps{ijk}'=\OO{2}, \quad \Ps{i}'=\OO{3} , \label{R=0_i_nu>=4} \\
			& & \Om{ij}'=\OO{2} , \nonumber 
		\eeqn
		{with the further restrictions $\WDS{ij}=\OOp{1-\nu}$ for $4\le\nu<5$ and $\WDS{ij}=\OO{4}$ for $\nu\ge5$.}

	\item 		
		
		\beqn
			& & \WD{ijkl}=\OO{2} , \quad \WDS{ij}=\OOp{1-\nu}, \quad \WD{}=\OO{\nu} , \quad \WDA{ij}=\OO{\nu}  \quad (n>5, \, \nu\ge4, \, \nu\neq n) , \nonumber  \\
			& & \Ps{ijk}'=\OO{2}, \quad \Ps{i}'=\OOp{1-\nu} , \\
			& & \Om{ij}'=\OO{2} . \nonumber 
		\eeqn

 \item 		
		
		\beqn
			& & \WD{ijkl}=\OO{2} , \qquad \WDS{ij}=\OOp{1-n}, \qquad \WDA{ij}=\OO{n}  \qquad (n>5, \, {\nu\ge n}) , \nonumber  \\
			& & \Ps{ijk}'=\OO{2}, \quad \Ps{i}'=\OOp{1-n} , \\
			& & \Om{ij}'=\OO{2} . \nonumber 
		\eeqn
				
	None of the above five cases can describe asymptotically flat spacetimes, cf.~\cite{GodRea12}.  In {cases A2--A5}, the leading term at infinity falls off as $1/r^2$ and it is of type II(abd). In {cases A3--A5}, $\bl$ can be a {multiple} WAND, cf. also the results of \cite{OrtPraPra_prep}. {Examples in case A5 are Robinson-Trautman Ricci-flat spacetimes \cite{PodOrt06}.}

	\end{enumerate}

	\item For any $n>5$, we have
	
		\beqn
			& & \WD{ijkl}=\OO{n/2} , \qquad \WD{}=\OO{\nu} , \qquad \WDA{ij}=\OO{\nu}  \qquad \left(n>5, \,\frac{n}{2}<\nu\le1+\frac{n}{2}\right) , \nonumber  \\
			& & \Ps{ijk}'=\OO{n/2} , \label{R=0_i_beta=-n/2} \\
			& & \Om{ij}'=\OOp{1-n/2} . \nonumber 
		\eeqn	
		
		Note that here $\bl$ cannot be a WAND.	The leading term at infinity falls off as $1/r^{n/2-1}$ and it is of type N. At order $1/r^{n/2}$ the type becomes II(acd) (as follows from section~\ref{subsubsec_R=0_bw0}).

		For $n=5$ the same behaviour applies if $3<\nu\le\frac{7}{2}$, while $\Ps{ijk}'=\OO{2}$ if $\frac{5}{2}<\nu\le3$ (the other terms being unchanged).		
		
		{If we take for b.w. +2 components $\nu=1+\frac{n}{2}$ and additionally {\em assume} that
			\be
			  \Om{ij}=\frac{\Om{ij}^{(n/2+1)}}{r^{n/2+1}}+\frac{\Om{ij}^{(n/2+2)}}{r^{n/2+2}}+\oop{-n/2-2} , \label{Omega_sublead}
			\ee
			then \Bi{4} with \Bi{5} show that the subleading term of $\Om{ij}'$ is of order $\OO{n/2}$, which with \eqref{R=0_i_beta=-n/2} implies the following peeling-off behaviour
				\be
					C_{abcd}=\frac{N_{abcd}}{r^{n/2-1}}+\frac{II_{abcd}}{r^{n/2}}+\oo{n/2} \qquad (n\ge5) .
					\label{peel_R=0_i}
				\ee
			This result is in agreement with the conclusions of \cite{GodRea12} for asymptotically flat spacetimes {(and extends it to asymptotics along twisting null geodesics)}. However, in order to obtain higher-order terms one would need to make further assumptions on how $\Om{ij}$ can be expanded, which goes beyond the analysis of the present paper (however, recall that it is precisely at a higher order in \eqref{peel_R=0_i} that \cite{GodRea12} found a qualitative difference between five and higher dimensions). In five dimensions, a permitted behaviour more general than \eqref{peel_R=0_i} is described in section~\ref{subsubsec_R=0_iii_n=5} below (it does not appear here because it belongs to case~(\ref{+1_3})).}

In view of \cite{GodRea12}, we conclude that the above behaviour \eqref{R=0_i_beta=-n/2} includes {\em radiative spacetimes} that are asymptotically flat in the Bondi definition \cite{TanTanShi10,TanKinShi11} (which is equivalent \cite{GodRea12} to the conformal definition \cite{HolIsh05,Ishibashi08} in even dimensions).

If one takes $\nu>1+\frac{n}{2}$ in \eqref{R=0_i_beta=-n/2}, this reduces to \eqref{R=0_generic} if $1+\frac{n}{2}<\nu\le n-1$, to \eqref{R=0_i_betac_=1-n} if $n-1<\nu\le n$, and to \eqref{R=0_ii_beta_c=1-n} if $\nu>n$.

	\item For $n>4$ the fall-off is 
	
	\beqn
			& & \WD{ijkl}=\OOp{1-n} , \qquad \WDA{ij}=\oop{1-n}  \qquad (n>4, \,\nu>n-1) , \nonumber  \\
			& & \Ps{ijk}'=\OOp{1-n} , \label{R=0_i_betac_=1-n} \\
			& & \Om{ij}'=\oop{2-n} , \nonumber 
		\eeqn	 
	with $\WDA{ij}=\OO{\nu}$ for $n-1<\nu<n$ and $\WDA{ij}=\OO{n}$ for $\nu\ge n$. Here $\bl$ can {become a multiple WAND, cf.~\cite{OrtPraPra09b,OrtPraPra_prep}.} This behaviour is compatible with the results of \cite{GodRea12} for asymptotically flat spacetimes, in the case of {\em vanishing radiation}. In particular, it includes asymptotically flat spacetimes for which $\bl$ is a multiple WAND \cite{OrtPraPra09b,OrtPraPra_prep}, such as Ricci flat Robinson-Trautman spacetimes \cite{PodOrt06} (e.g., Schwarzschild-Tangherlini black holes), {and Kerr-Schild spacetimes \cite{OrtPraPra09} with a non-degenerate Kerr-Schild vector\footnote{For these one finds $\Om{ij}'=\OOp{1-n}$. Note that in order to explicitly verify this using the general expressions given in \cite{OrtPraPra09} one should recall to enforce the vacuum equation $R_{11}=0$, cf.~\cite{Krssak09}. The same comment applies to the (A)dS Kerr-Schild spacetimes \cite{MalPra11} mentioned in section \ref{subsubsec_R_summary}.} (e.g., Myers-Perry black holes).}

		For $n=4$ we have instead
		
		\beqn
			& & \WD{ijkl}=\OO{3} , \qquad \WDA{ij}=\OO{3}  \qquad (n=4, \,\nu>3) , \nonumber  \\
			& & \Ps{ijk}'=\OO{2} , \label{R=0_i_n=4} \\
			& & \Om{ij}'=\OO{1} , \nonumber 
		\eeqn	
		where the leading $1/r$ term is of type N. However, this is not the ``standard'' four-dimensional peeling behaviour, which would require the stronger condition $\nu=5$ \cite{NP}. Generalized peeling properties under asymptotic conditions weaker than those of \cite{NP} have been already studied in four dimensions, e.g., in \cite{CouTor72,ChrKla93,ChrMacSing95,Kroon98}. We note that the assumption made in this paper that 
		leading-order terms of Weyl components are power-like is in fact generically too restrictive in those cases (for example, for $\nu=4$ the natural framework to consider is that of polyhomogenous expansions \cite{Kroon98}). Similar comments will apply to \eqref{R=0_n=4_nu>2} below.
		\newline

\end{enumerate}

\subsubsection{Special subcase $\beta_c=\beta=-\nu$}

\label{subsubsec_R=0_i_special}

In addition, there is the case $\beta_c=\beta=-\nu$ (briefly mentioned in section~\ref{subsubsec_R=0_bw0} above but not explicitly studied in sections~\ref{subsubsec_R=0_bw-1} and \ref{subsubsec_R=0_bw-2}), for which one easily arrives {for $n>4$} at (note that \eqref{R=0_Ni1} still applies here) 
\beqn
	& & \WD{ijkl}=\OO{\nu}, \qquad \WDA{ij}=\OO{\nu},  \qquad\qquad\qquad\qquad\qquad\qquad {(n>4)} \nonumber \\
	& & \Ps{ijk}'=\OO{2} \quad\mbox{if } 2<\nu\le3, \qquad \Ps{ijk}'=\OO{\nu} \quad\mbox{if } \nu>3, \label{R=0_generic} \\
	& & \Om{ij}'=\oop{1-\nu}  \quad\mbox{if } \nu\neq\frac{n}{2}, \qquad \Om{ij}'=\OOp{1-n/2}  \quad\mbox{if } \nu=\frac{n}{2}, \nonumber 
\eeqn
{with $X^{A0}\Om{ij,A}^{(\nu)}+(\nu-2)l_{11}\Om{ij}^{(\nu)}+2\Om{s(j}^{(\nu)}\m{s}{i)}{1}=0$.}
$\bl$ cannot be a WAND. 
{The above conditions on $\Om{ij}'$ have been obtained by using \Bi{4} and the trace of \Bi{13}.}

For $n=4$ one finds instead 
\beqn
	& & \WD{ijkl}=\OO{\nu}, \qquad \WDA{ij}=\OO{\nu},  \qquad\qquad {(n=4, \, \nu>2)} \label{R=0_n=4_nu>2} \nonumber \\
	& & \Ps{ijk}'=\OO{2} , \\
	& & \Om{ij}'=\OO{1} , \nonumber 
\eeqn
which is asymptotically of type N. {For $\nu>4$}, this is a subcase of \eqref{R=0_n=4_ii} having $\WD{ijkl}^{(3)}=0$, $\Phi_{ij}^{A(3)}=0$ and $\Ps{ijk}^{(4)}=0$.

\subsection{Case (ii): $\Ps{ijk}=\OO{n}$, $\Om{ij}=\oo{n}$}

\label{subsec_R=0_ii}

The behaviour of the Ricci rotation coefficients and derivative operators is the same as in case (\ref{+1_gener}) (in particular, \eqref{N_R=0}, \eqref{L11_R=0}, \eqref{Mij1}, \eqref{U_R=0} and \eqref{R=0_Ni1} still apply).

\subsubsection{Case $\beta_c=-2$, $n>5$}

All the following cases can occur only for $n>5$.

\beqn
	& & \Om{ij}=\oo{n} , \nonumber \\	
	& & \Ps{ijk}=\OO{n} , \nonumber \\
	& & \WD{ijkl}=\OO{2}, \qquad \WDS{ij}=\OO{4} , \qquad \WDA{ij}=\OO{3} ,  \label{R=0_ii_betac=-2} \\
	& & \Ps{ijk}'=\OO{2} , \qquad \Ps{i}'=\OO{3} , \nonumber \\
	& & {\Om{ij}'=\OO{2}} , \nonumber 
\eeqn
with $(n-4)\Phi_{ij}^{A(3)}=\WD{ikjl}^{(2)}b_{[kl]}$ {and $\WD{ikjl}^{(2)}b_{(kl)}=0$}. Here $\bl$ can be a single WAND. For $\Ps{ijk}^{(n)}=0$ this case reduces to \eqref{R=0_i_nu>=4} (with $\nu>n$).

If $\WD{ikjl}^{(2)}b_{[kl]}=0$ {(in particular, if $\bl$ is twist free)} the following subcase arises:

\beqn
	& & \Om{ij}=\oo{n} , \nonumber \\
	& & \Ps{ijk}=\OO{n} , \nonumber \\
	& & \WD{ijkl}=\OO{2} , \qquad \WDS{ij}=\OOp{1-n} , \qquad \WDA{ij}=\OO{n} ,  \\
	& & \Ps{ijk}'=\OO{2} , \qquad \Ps{i}'=\OOp{1-n}   ,\nonumber \\
	& & {\Om{ij}'=\OO{2}} , \nonumber 
\eeqn
{with $(2-n)\Phi^{S(n-1)}_{ij}+\Phi^{(n-1)}\delta_{ij}=0$.} 

As a {further} ``subcase'', if {$\Phi_{ij}^{S(n-1)}=0$} we obtain, {depending on the value of $\nu$,}
\beqn
	& & \Om{ij}=\OO{\nu} \qquad (n<\nu\le n+1) , \nonumber \\
	& & \Ps{ijk}=\OO{n} , \nonumber \\
	& & \WD{ijkl}=\OO{2} , \qquad \WDS{ij}=\OOp{1-\nu}, \qquad \WD{}=\OO{n} , \qquad \WDA{ij}=\OO{n} ,  \\
	& & \Ps{ijk}'=\OO{2} , \qquad \Ps{i}'=\OOp{1-\nu} ,\nonumber \\
	& & {\Om{ij}'=\OO{2}} , \nonumber 
\eeqn
{or
\beqn
	& & \Om{ij}=\OO{\nu} \qquad (\nu>n+1) , \nonumber \\
	& & \Ps{ijk}=\OO{n} , \nonumber \\
	& & \WD{ijkl}=\OO{2} , \qquad \WDS{ij}=\OO{n}, \qquad \WDA{ij}=\OO{n} ,  \\
	& & \Ps{ijk}'=\OO{2} , \qquad {\Ps{i}'=\OOp{1-n}} ,\nonumber \quad  \\
	& & {\Om{ij}'=\OO{2}} . \nonumber 
\eeqn
}

In all the above cases $\Ps{ijk}'^{(2)}=-\WD{isjk}^{(2)}l_{s1}$ and $\Om{ij}'^{(2)}=-\Ps{(ij)k}'^{(2)}l_{k1}=\WD{isjk}^{(2)}l_{s1}l_{k1}$. The asymptotically leading term is of type II(abd) but it reduces to type D(abd) if a particular frame with $l_{i1}=0$  is employed {cf. the comments at the end of section~\ref{subsec_Rc0}.} The terms $\WD{ijkl}=\OO{2}$ violate the asymptotically flat conditions \cite{GodRea12}.

\subsubsection{Case $\beta_c<-2$, {$n>4$}}

If $\WD{ijkl}^{(2)}=0$ one is left with

\beqn
	& & \Om{ij}=\oo{n} , \nonumber \\
	& & \Ps{ijk}=\OO{n} , \nonumber \\
  & & \WD{ijkl}=\OOp{1-n} , \qquad \WDA{ij}=\OO{n} ,  \label{R=0_ii_beta_c=1-n} \ \\	
	& & \Ps{ijk}'=\OOp{1-n} , \qquad \Ps{i}'=\OOp{1-n}   ,\nonumber \\
	& & \Om{ij}'=\oop{2-n} , \nonumber 
\eeqn
with $(n-2)(n-3)\WD{ijkl}^{(n-1)}=4\Phi^{(n-1)}\delta_{j[k}\delta_{m]i}$, {$(n-3)\Ps{ijk}^{(n-1)}=2\Ps{[j}^{(n-1)}\delta_{k]i}$, $(n-2)\Ps{i}^{(n-1)}=-(n-1)\WD{}^{(n-1)}l_{i1}$,} and where $\bl$ can be a single WAND. This behaviour is compatible with the results of \cite{GodRea12} for asymptotically flat spacetimes, in the case of vanishing radiation.
For $\Ps{ijk}^{(n)}=0$, this case reduces to \eqref{R=0_i_betac_=1-n} (with $\nu>n$).  

If $\Phi^{(n-1)}=0$ this reduces to 
\beqn
	& & \Om{ij}=\oo{n} , \nonumber \\
	& & \Ps{ijk}=\OO{n} , \nonumber \\
  & & \WD{ijkl}=\OO{n} , \qquad \WDA{ij}=\OO{1-n} ,  \\	
	& & \Ps{ijk}'=\OO{n} ,  \nonumber \\
	& & \Om{ij}'=\OOp{1-n} . \nonumber 
\eeqn
The asymptotically leading term is of type N.

\subsubsection{Case $n=4$}

\beqn
	& & \Om{ij}=\OO{\nu} \qquad (\nu>4) , \nonumber \\
	& & \Ps{ijk}=\OO{4} , \nonumber \\
	& & \WD{ijkl}=\OO{3}, \qquad \WDA{ij}=\OO{3},  \label{R=0_n=4_ii} \\
	& & \Ps{ijk}'={\OO{2}}  ,\nonumber \\
	& & \Om{ij}'={\OO{1}} . \nonumber 
\eeqn

The above behaviour agrees with the well-known results of \cite{NP} (where it was assumed $\nu=5$). {For $\Ps{ijk}^{(4)}=0$ this case reduces to \eqref{R=0_i_n=4} (with $\nu>4$).} See \cite{NewUnt62} for results also at the subleading order.

\subsection{Case (iii): $\Ps{ijk}=\OO{3}$, $\Om{ij}=\oo{3}$ ($n>4$)}

\label{subsec_R=0_iii}

Again the behaviour of the Ricci rotation coefficients and derivative operators is the same as in case (\ref{+1_gener}).

\subsubsection{Case $n>5$}

In more than five dimensions we generically have $\beta_c=-2$, giving rise to 

\beqn
	& & \Om{ij}=\OO{\nu} \qquad (\nu>3) , \nonumber \\
	& & \Ps{ijk}=\OO{3} , \qquad \Ps{i}=\oo{3} , \nonumber \\
	& & \WD{ijkl}=\OO{2}, \qquad \WDS{ij}=\OO{3} , \qquad \WD{}=\oo{3} , \qquad \WDA{ij}=\OO{3} ,  \\
	& & \Ps{ijk}'=\OO{2} , \qquad \Ps{i}'=\OO{3}  ,\nonumber \\
	& & {\Om{ij}'=\OO{2}} , \nonumber 
\eeqn
where $\Ps{i}=\OO{\nu}$, $\WD{}=\OO{\nu}$ for $3<\nu\le 4$ while $\Ps{i}=\OO{4}$, $\WD{}=\OO{4}$ for $\nu>4$ and

\beqn
 & & (n-4)\Phi_{ki}^{A(3)}=\WD{klij}^{(2)}b_{[lj]}+\xi^{A0}_j\Ps{[ki]j,A}^{(3)}+2l_{1j}\Ps{[ki]j}^{(3)}+\Ps{[ki]l}^{(3)}\m{l}{j}{j}+\Ps{jl[k}^{(3)}\m{l}{i]}{j} , \nonumber \\
 & & (n-6)\Phi_{ki}^{S(3)}=\WD{klij}^{(2)}b_{(lj)}+\xi^{A0}_j\Ps{(ki)j,A}^{(3)}+2l_{1j}\Ps{(ki)j}^{(3)}+\Ps{(ki)l}^{(3)}\m{l}{j}{j}+(2\Ps{l(k|j}^{(3)}+\Ps{jl(k|}^{(3)})\m{l}{|i)}{j} . \nonumber 
\eeqn
Here $\bl$ can be a single WAND and the asymptotically leading term is of type II(abd). {For $\Ps{ijk}^{(3)}=0$, this case reduces for $3<\nu<4$ to \eqref{R=0_i_3<nu<4} (with $\nu>n$), for $4\le\nu\le n$ to \eqref{R=0_i_nu>=4} and for $\nu>n$ to \eqref{R=0_ii_betac=-2}. }

A subcase with $\WD{ijkl}^{(2)}=0$ is also possible, giving
\beqn
	& & \Om{ij}=\OO{\nu} \qquad (\nu>3) , \nonumber \\
	& & \Ps{ijk}=\OO{3} , \qquad \Ps{i}=\oo{3} , \nonumber \\
	& & \WD{ijkl}=\OO{3}, \qquad \WD{}=\oo{3} , \qquad \WDA{ij}=\OO{3} ,  \label{R=0_iii_subcase} \\
	& & \Ps{ijk}'=\OO{2} , \qquad \Ps{i}'=\OO{3}   ,\nonumber \\
	& & \Om{ij}'=\OO{2} , \nonumber 
\eeqn
with the same behaviour as above for $\Ps{i}$ and $\WD{}$. In this case the leading term at infinity is of type III(a).

 {Neither of the above behaviours can represent asymptotically flat spacetimes since the fall-off of the Weyl tensor is too slow \cite{GodRea12}.}

\subsubsection{Case $n=5$}

\label{subsubsec_R=0_iii_n=5}

In five dimensions, we generically have
\beqn
	& & \Om{ij}=\OO{\nu} \qquad (3<\nu\le\textstyle{\frac{7}{2}}) , \nonumber \\
	& & \Ps{ijk}=\OO{3} , \qquad \Ps{i}=\OO{\nu} , \nonumber \\
	& & \WD{ijkl}=\OO{5/2}, \qquad \WD{}={\OO{\nu}} , \qquad \WDA{ij}=\OO{3} ,  \label{R=0_iii_n=5} \\
	& & \Ps{ijk}'=\OO{2} , \qquad \Ps{i}'=\OO{3}    ,\nonumber \\
	& & \Om{ij}'=\OO{3/2} , \nonumber 
\eeqn
with $\Phi_{ki}^{A(3)}=\xi^{A0}_j\Ps{[ki]j,A}^{(3)}+2l_{1j}\Ps{[ki]j}^{(3)}+\Ps{[ki]l}^{(3)}\m{l}{j}{j}+\Ps{jl[k}^{(3)}\m{l}{i]}{j}$, $\Ps{ijk}'^{(2)}$ can be expressed in terms of $\Ps{ijk}^{(3)}$ using \Bi{6} and $\Om{ij}'^{(3/2)}=-5l_{11}\Phi_{ij}^{S(5/2)}-2X^{A0}\Phi_{ij,A}^{S(5/2)}-4\Phi_{s(j}^{S(5/2)}\m{s}{i)}{1}$. If $\nu=7/2$ this can be rewritten using $3\Phi_{ij}^{S(5/2)}=-2X^{A0}\Om{ij,A}^{(7/2)}-3l_{11}\Om{ij}^{(7/2)}-4\Om{s(j}^{(7/2)}\m{s}{i)}{1}$. Recalling the comments following \eqref{R=0_i_beta=-n/2}, one finds that the same behaviour {\eqref{R=0_iii_n=5}} holds in fact for the full range  $\frac{5}{2}<\nu\le\frac{7}{2}$ (unless $\Ps{ijk}^{(3)}=0$). In all cases here $\bl$ cannot be a WAND, and the asymptotically leading term is of type N.	

{Note an important difference with the behaviour \eqref{R=0_i_beta=-n/2} with $n=5$: after the leading type N term, the subleading term in \eqref{R=0_iii_n=5} is of type III(a) (it was of type II(acd) in \eqref{R=0_i_beta=-n/2}). If we assume for $\Om{ij}$ a fall-off as in \eqref{Omega_sublead}, this shows that the subleading term of $\Om{ij}'$ is of order $\OO{2}$, thus leading to the qualitatively different peeling-off behaviour
\be
	C_{abcd}=\frac{N_{abcd}}{r^{3/2}}+\frac{III_{abcd}}{r^{2}}+\oo{2} \qquad (n=5) .
	\label{peel_R=0_ii_n=5}
\ee
However, according to \cite{GodRea12} this behaviour is not permitted in asymptotically flat spacetimes. For the latter one thus concludes that $\Ps{ijk}^{(3)}=0$ (in which case \eqref{R=0_iii_n=5} reduces to \eqref{R=0_i_beta=-n/2} with $n=5$) is a necessary boundary condition in five dimensions. This is perhaps not surprising since $\Ps{ijk}^{(3)}=0$ already in four dimensions (where $\Ps{ijk}=\OO{4}$ \cite{NP}, cf. also \eqref{R=0_n=4_ii} above).}

If $\nu>7/2$ the asymptotic behaviour is described by \eqref{R=0_iii_subcase} (in which cases $\bl$ can be a single WAND). 

{All the above results for the case $\tilde R=0$ are summarized in table~\ref{tab_R=0}}.

\begin{footnotesize}
\begin{sidewaystable}[!htbp]
\footnotesize
 \begin{center}
   \begin{tabular}{|c ccccccccc|}
    \hline case & $\Omega_{ij}$ & $\Psi_{ijk}$, $\Psi_i$ & $\Phi_{ijkl}$ & $\Phi^S_{ij}$, $\Phi$ & $\Phi^A_{ij}$& $\Psi'_{ijk}$, $\Psi'_i$ & $\Omega'_{ij}$ & restrictions & comments \\ \hline

(i) A1 & {$r^{-\nu}$} & {$r^{-\nu}$} & {$r^{-2}$} &${o}(r^{-2})$  & ${o}(r^{-2})$& $r^{-2}$& {$r^{\sigma}$} & $n>5$, $2<\nu\leq 3$, $-2\leq \sigma<-1$ & {$\bl$ not a WAND} \\ [1mm]
(i) A2 &  {$r^{-\nu}$} & {$r^{-\nu}$} & {$r^{-2}$} & $r^{-3}$, $r^{-\nu}$&  $r^{-3}$ & $r^{-2}$, $r^{-3}$ &  $r^{-2}$ & $n>5$, $3<\nu<4$& {$\bl$ not a WAND}  \\ [1mm]

(i) A3  &  $r^{-\nu}$  &{$r^{-\nu}$} &  $r^{-2}$ & {$r^{1-\nu}$}, {$r^{-4}$} & {$r^{-3}$}  & {$r^{-2}$,  $r^{-3}$} & $r^{-2}$ &  $n>5$, $4\leq\nu<5$ 
 & {$\bl$ not a WAND}  \\[1mm]
 & {$r^{-\nu}$} & {$r^{-\nu}$} &  {$r^{-2}$}  &  $r^{-4}$ & {$r^{-3}$} & {$r^{-2}$,  $r^{-3}$}  & {$r^{-2}$}  & $n>5$, $\nu\geq 5$& \\[1mm]
(i) A4 & $r^{-\nu}$ & {$r^{-\nu}$} & $r^{-2}$&  $r^{1-\nu}$,  $r^{-\nu}$  & $r^{-\nu}$ & $r^{-2}$,  $r^{1-\nu}$  & $r^{-2}$ & $n>5$, $\nu\geq 4$, $\nu\not= n$ &
 \\ [1mm]
(i) A5 & {$r^{-\nu}$} & {$r^{-\nu}$} & $r^{-2}$&  $r^{1-n}$  & $r^{-n}$ & $r^{-2}$,  $r^{1-n}$  & $r^{-2}$ & $n>5$, $\nu\geq n$ & {includes RT} \\ [1mm]\hline
(i) (B) & $r^{-\nu}$ & {$r^{-\nu}$}&$r^{-n/2}$ &  $r^{-n/2}$, $r^{-\nu}$ &$r^{-\nu}$&$r^{-n/2}$ &{$r^{1-n/2}$} &{$n>5$} and $n/2<\nu\leq n/2+1$  & {radiation}, {$\bl$ not a WAND}\\[1mm]
&  & & &   & & & & or $n=5$ and $3<\nu\leq 7/2$ & {$\bl$ not a WAND}\\[1mm]
 & $r^{-\nu}$ & {$r^{-\nu}$} &{$r^{-5/2}$} &  {$r^{-5/2}$}, $r^{-\nu}$ &$r^{-\nu}$&$r^{-2}$ &{$r^{-3/2}$} & $n=5$ and $5/2<\nu\leq 3$ & {$\bl$ not a WAND} \\[1mm]\hline
(i) (C) 
 & {$r^{-\nu}$} & {$r^{-\nu}$} &${r^{1-n}}$ &   &${o}({r^{1-n}})$&${r^{1-n}}$ &{${o}(r^{2-n})$} &     $\nu>n-1$ & {includes RT, KS}\\[1mm]
%
%
%
%
%
 & {$r^{-\nu}$}  & {$r^{-\nu}$} &  $r^{-3}$  & & $r^{-3}$ & $r^{-2}$ & $r^{-1}$ &$n=4$, {$\nu> 3$}&  \\ [1mm] 
\hline\hline

(ii) & {${o}(r^{-n})$} & {$r^{-n}$} & $r^{-2}$  &$r^{-4}$  &$r^{-3}$  &$r^{-2}$ , $r^{-3}$  &$r^{-2}$  & {$n>5$} & \\[1mm]
& {${o}(r^{-n})$} & {$r^{-n}$} & $r^{-2}$  &${r^{1-n}}$  &$r^{-n}$  &$r^{-2}$ , ${r^{1-n}}$  &$r^{-2}$  & {$n>5$} & \\[1mm]
& {$r^{-\nu}$} & {$r^{-n}$}  & $r^{-2}$  & {$r^{1-\nu}$},  $r^{-n}$ &$r^{-n}$  &$r^{-2}$ , {$r^{1-\nu}$} &$r^{-2}$  & ${5<}n<\nu\leq n+1$ & {$\bl$ not a WAND} \\[1mm]
& {$r^{-\nu}$} & {$r^{-n}$} & $r^{-2}$  &$r^{-n}$ &$r^{-n}$  &$r^{-2}$ , ${r^{1-n}}$  &$r^{-2}$  &{$n>5$}, $\nu> n+1$ & \\[1mm]\hline
& {${o}(r^{-n})$} & {$r^{-n}$} & ${r^{1-n}}$ & &$r^{-n}$  &${r^{1-n}}$  &{${o}(r^{2-n})$}  & & \\[1mm]
& {${o}(r^{-n})$} & {$r^{-n}$} & $r^{-n}$ & &$r^{-n-1}$  &$r^{-n}$  &${r^{1-n}}$  & & \\[1mm]
& {$r^{-\nu}$} &$r^{-4}$ & $r^{-3}$ & &$r^{-3}$  &$r^{-2}$  &$r^{-1}$  & $n=4$, $\nu>4$ & \\[1mm]\hline\hline

(iii) & {$r^{-\nu}$} &$r^{-3}$, $r^{-\nu}$ & $r^{-2}$  &$r^{-3}$, $r^{-\nu}$  &$r^{-3}$  &$r^{-2}$ , $r^{-3}$  &$r^{-2}$  & {$n>5$}, $3<\nu\leq 4$& {$\bl$ not a WAND} \\[1mm]
& {$r^{-\nu}$} & $r^{-3}$, $r^{-4}$ & $r^{-2}$  &$r^{-3}$,  $r^{-4}$ &$r^{-3}$  &$r^{-2}$ , $r^{-3}$  &$r^{-2}$  & {$n>5$}, $\nu>4$ & \\[1mm]
& {$r^{-\nu}$}  & $r^{-3}$, ${o}(r^{-3})$& $r^{-3}$  &$r^{-3}$, ${o}(r^{-3})$ &$r^{-3}$  &$r^{-2}$ , $r^{-3}$  &$r^{-2}$  & {$n>5$} & \\[1mm]

&  & & &   & & & & {or $n=5$ and $\nu>7/2$} & \\[1mm]

& {$r^{-\nu}$}  &$r^{-3}$, $r^{-\nu}$ & $r^{-5/2}$ & $r^{-5/2}$, $r^{-\nu}$  &$r^{-3}$&$r^{-2}$, $r^{-3}$  &$r^{-3/2}$  &$n=5$, $5/2<\nu\leq 7/2$ & {$\bl$ not a WAND}\\[1mm]

	 \\[1mm] \hline           
  \end{tabular}
  \caption{Fall-off behaviour of the Weyl tensor for Ricci-flat spacetimes ($\tilde R=0$), listing the cases summarized in sections \ref{subsubsec_R=0_summary}, \ref{subsec_R=0_ii}, \ref{subsec_R=0_iii}. The conventions explained in Table~\ref{tab_R} apply also here. {Note that, for brevity, we have not included here the very special subcase of section~\ref{subsubsec_R=0_i_special}.}\label{tab_R=0}}
 \end{center}
\end{sidewaystable}
\end{footnotesize}

\section*{Acknowledgments}

The authors acknowledge support from research plan {RVO: 67985840} and research grant GA\v CR 13-10042S.


\end{document}